\documentclass[fleqn, 10pt,twocolumn] {article}

\usepackage{times}
\usepackage{amssymb}
\usepackage{amsmath}
\usepackage{amsfonts}
\usepackage{graphicx}
\usepackage{subfigure}
\usepackage{wrapfig}
\usepackage{changepage}
\usepackage{comment}
\usepackage{url}
\usepackage{eucal}
\usepackage[utf8]{inputenc}
\usepackage{caption}
\usepackage{cancel}
\usepackage{titlesec}
\usepackage{indentfirst}

\let\oldthefootnote\thefootnote

\DeclareCaptionLabelFormat{nocounter}{#2}

\titleformat{\section}
  {\normalfont\fontsize{12}{15}\centering \selectfont \bfseries \MakeUppercase} % Font style and size
  {\thesection.} % Section number formatting
  {1em} % Space between number and title
  {} % Title formatting

\usepackage[left=20mm, right=20mm, top=25.4mm, bottom=25.4mm, headsep=7mm, headheight=12pt]{geometry}

\title{Distributed and Decentralized Control and Task Allocation\\ for Flexible Swarms}
\author{Yigal Koifman, Ariel Barel and Alfred M. Bruckstein \textsuperscript{1}% <-this stops a space
}

\begin{document}
\twocolumn[
  \begin{@twocolumnfalse}
    \maketitle
    \begin{abstract}
        This paper introduces a novel bio-mimetic approach for distributed control of robotic swarms, inspired by the collective behaviors of swarms in nature such as schools of fish and flocks of birds. The agents are assumed to have limited sensory perception, lack memory, be identical, anonymous, and operate without inter-agent explicit communication. 
        Despite these limitations, we demonstrate that collaborative exploration and task allocation can be executed by applying simple local rules of interactions between the agents.
        A comprehensive model comprised of agent, formation, and swarm layers is proposed in this paper, where each layer performs a specific function in shaping the swarm's collective behavior, thereby contributing to the emergence of the anticipated behaviors.
        We consider four principles combined in the design of the distributed control process: Cohesiveness, Flexibility, Attraction-Repulsion, and Peristaltic Motion. We design the control algorithms as reactive behaviour that enables the swarm to maintain connectivity, adapt to dynamic environments, spread out and cover a region with a size determined by the number of agents, and respond to various local task requirements.
        We explore some simple broadcast control-based steering methods, that result in inducing “anonymous ad-hoc leaders” among the agents, capable of guiding the swarm towards yet unexplored regions with further tasks.
        Our analysis is complemented by simulations, validating the efficacy of our algorithms. The experiments with various scenarios showcase the swarm's capability to self-organize and perform tasks effectively under the proposed framework.
        The possible implementations include domains that necessitate emergent coordination and control in multi-agent systems, without the need for advanced individual abilities or direct communication.
    \end{abstract}
    \vspace{1cm}
  \end{@twocolumnfalse}
]

%-----------------------------------------------------------------------
 
 % \noindent\textbf{Keywords: }\keywords{Bio-inspired agents, autonomous swarm, swarm robotics, distributed and decentralised control, task allocation, multi-agent systems, swarm steering.}

\renewcommand{\thefootnote}{}
\footnotetext{ \textsuperscript{1}The authors are with the Department of Computer Science, Technion Israel Institute of Technology, Haifa, Israel. E-mails: {\tt\small \{igal.k, freddy\}@cs.technion.ac.il, arielba@technion.ac.il}.}
\let\thefootnote\oldthefootnote

\section{Introduction}
\begin{figure}[ht]
\begin{center}
\includegraphics[width=8cm]{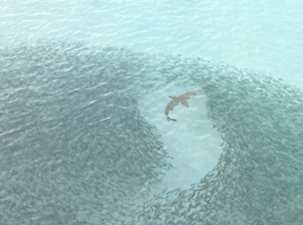}
\caption{A school of fish: the school formation changes to increase survivability, yet keeps cohesiveness, in response to the predator's presence. Photograph by the first author, taken in the Maldives in $2022$.}
\label{fig:shoal of fish}
\end{center}
\end{figure}

One of nature’s most captivating phenomena is the swarming behaviour \cite{nationalgeographic2021birds}. A swarm refers to a large group of individual entities or agents that operate together in a cohesive and coordinated manner, carrying out activities such as aggregating and patrolling a region, foraging, and migration. These can be observed in flocks of birds, schools of fish, herds of bisons, and bee, termite, and ant colonies, among other species. Swarming behavior is driven by common objectives among group members, such as improving survivability by the ``confusion effect” \cite{hogan2017confusion} and by the ``oddity effect” \cite{landeau1986oddity,mattila2021giant}, as illustrated in Fig. \ref{fig:shoal of fish}, 
reproduction effectiveness, energy conservation \cite{portugal2014upwash}, searching or hunting for food \cite {bruckstein1991ant}, etc.

Swarming behaviours offer many advantages, such as leaderless self-organization, autonomously emerging collaborative activities, and adaptability to changes in the environment, executed in a decentralized manner, where each individual makes its own decisions based on local information.

\subsection {Overview}
Throughout human history, inquisitive minds have sought to comprehend the social behavior of animals. This pursuit, driven by inherent curiosity and potential applications, prompted interest in the collective behaviors of swarms and herds of animals \cite{schranz2020swarm,okubo1986dynamical}.

A prime example of such social behavior can be observed in an ant colony \cite{dorigo2000ant}. Ants operate with limited sensory range and lack a global view of the swarm and the environment. Nevertheless, ants collaborate efficiently, following paths agreed upon by the colony, even though the paths may not necessarily be globally optimal. Numerous research papers have been published on collective behavior in nature, several of which are referenced in this study \cite{sumpter2006principles,ballerini2008empirical,hildenbrandt2010self}.

In an earlier study, Amir et al. \cite{amir2023multi} demonstrate that simple agents following local behavioural rules can position themselves according to task requirements when placed in a \textit{bounded area}, satisfying task demands.
These rules, which allow agents to overcome their limited sensory range, are based on the assumptions that they are confined in closed areas, and that the swarm comprises significantly more agents than necessary to fulfill the overall requirements, consequently the swarm dispersion is prevented.
\subsection{The Research Goal}
At the core of our current research lies the trade-off between limited sensory range and efficient emergent collaboration.
The difficulties faced when trying to imitate the emergence of complex self-organizing behaviors stem mostly from the swarm size and the complexity of the mutual interactions among the swarm’s members and the environment. It is particularly hard to control and predict the emerging self-organizing behavior in large-scale swarms.

The outcome of our work is a set of individual, localized autonomous behavioral rules that promote large-scale swarms to act in collaborative behaviour while being steered toward a desired area and to meet tasks requirements in a \textit{desired area}, without having to control each agent individually.

Our approach proposes a multi-layer model of behaviour for multi-agent systems of autonomous agents, based on certain basic assumptions about their characteristics.

\subsection {Assumptions and Notation}
The model is based on the paradigms that agents are identical and indistinguishable, lack explicit communication capabilities, do not share a common geometric frame of reference, are oblivious, and have limited sensory range. These limitations exemplify our aim to create a model that mimics the actions of simple organisms.
The agents are assumed to operate in the plane $\mathbb{R}$$^{2}$
 that contains $K$ target tasks which are a priori unknown to the agents, denoted as
$T \triangleq \{(t_i)\}_{i=1, 2, 3,\ldots	,K}$ and represented as markers. Each task induces a spatial demand function $d_i(x,y)$ that signifies the needed sensory coverage at a given location $(x,y)$ within the plane.

We denote agents as $A \triangleq \{(a_i)\}_{i=1, 2, 3,\ldots	,N}$, their positions at time $t$ as $\vec{P}(t) \triangleq \{\vec{p}_i(t)\}_{i=1, 2, 3, \ldots,	N}$ where $\vec{p}_i(t) \triangleq \{(x_i,y_i)^T\}_{i=1, 2, 3, \ldots	,N}$, 
and the tasks' positions as $\vec{C} \triangleq \{\vec{c_k}\}_{k=1, 2, 3,\ldots,K}$, where $\vec{c_k} \triangleq (x_k,y_k)^T$. Hence the Euclidean distance between agents $a_i$ and $a_j$ is $\parallel \vec{p}_i - \vec{p}_j \parallel$ and the Euclidean distance between agent $a_i$ and task $t_k$ is $\parallel \vec{p}_i - \vec{c}_k \parallel$.

\section {A Multi-Layered Approach for Task Allocation}
\begin{figure}[!ht]
\begin{center}
\includegraphics[width=7.5cm]{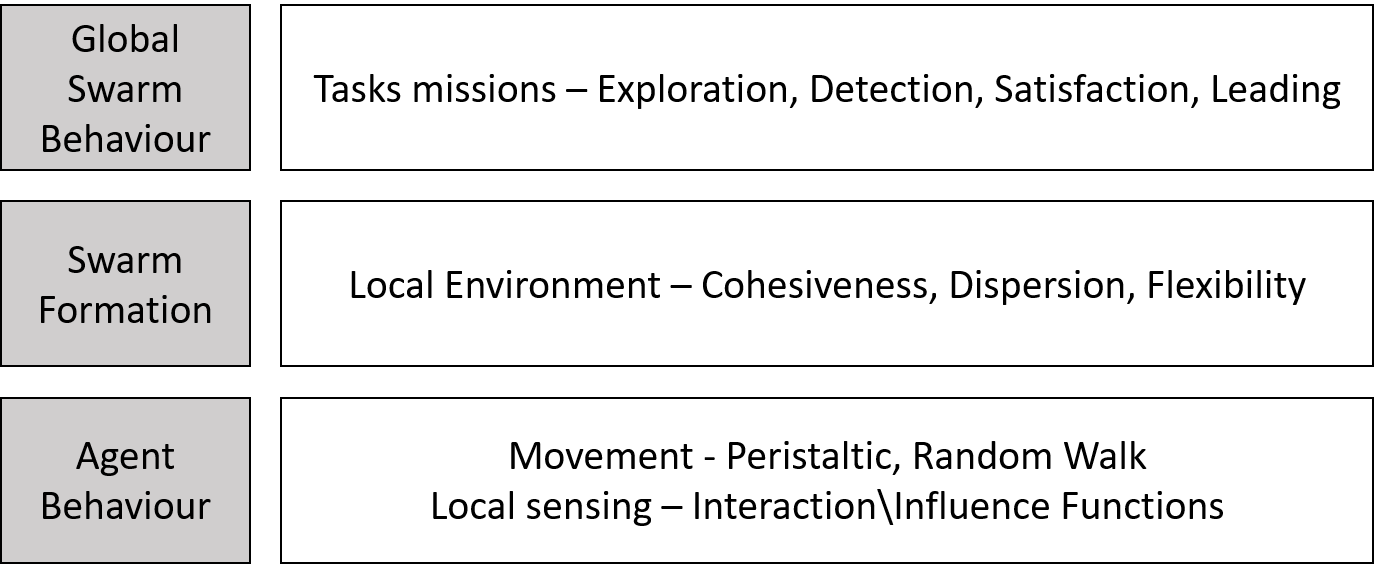}
\caption{Description of the swarm multi-layered behavior that emerges from the local dynamics rules}
\label{fig:TheModel}
\end{center}
\end{figure}

We present a multi-layered behavior approach for emergent task allocation, as illustrated in Fig. \ref{fig:TheModel}. It comprises three behavior layers: an agent behavior layer, a swarm formation behavior layer, and a global swarm behavior layer.
The agent layer comprises rules that derive from local sensing and individual maneuvering. The formation layer is composed of rules based on the agent interactions that impact the swarm constellation, hence influencing the swarm as a cohesive system. Lastly, the global swarm behavior layer defines the autonomously emerging behaviors of exploration, motion, and task allocation.\\ 

With this model in mind, we next present the local rules of motion:

\subsection{Local Motion Rules}
To achieve natural behavior that mimics nature, we suggest several distributed motion rules that influence the mutual dynamics of agents. In the sequel, we elaborate on these rules, and demonstrate how their combination yields the desired behavior of the swarm:

\begin{itemize}
\item{\textbf{Swarm Cohesiveness} (Sec \ref{sec:swarm Cohesiveness})}: A fundamental requirement is that the swarm stays cohesive, that is, to persist as a connected entity and prevent fractioning into multiple swarms.

\item{\textbf{Swarm Flexibility} (Sec \ref{sec: Swarm Flexibility})}: Swarms should be adaptable while also maintain connections among agents that uphold the swarm cohesiveness. This requires that enough pairs of agents maintain a mutual distance no greater than their visibility range
so that the ``visibility" graph of the swarm remains connected.
\item{\textbf{Attraction-Repulsion Interaction} (Sec \ref{sec: Attraction Repulsion Interaction}}): This rule compels agents to targets, and propels agents to distance themselves to be able to expand the coverage area of the swarm.

\item{\textbf{Peristaltic Mechanism} (Sec \ref{sec:The Peristaltic Effect})}: This mechanism provides random steps made to enable enhanced mobility, reducing the rigidity of the constellation. This also helps enhance the swarm's exploration capabilities.

\item{\textbf{Swarm Steering} (Sec \ref{sec:Steering and Guiding the Swarm})}: This feature enables an external observer or certain sub-groups of agents to control the global motion of the swarm. Here we proceed on the basis of two possible assumptions. First, we consider an external observer, who can detect the swarm's and tasks' position and aim to steer the swarm toward the task zone by broadcasting a control direction to the entire swarm. This control signal is detected with some probability by the agent. The agent that detect the signal becomes ad-hoc \textbf{anonymous leader}. Alternatively, one may assume that some agents, can acquire information about the tasks relative location to their current location, and direct other agents to move toward the task area, as anonymous leaders themselves.
\end{itemize}

By consolidating all of the aforementioned parameters, an enhanced and more comprehensive exploration capabilities are achieved. Let us elaborate on these local motion rules:

\subsubsection{Swarm Cohesiveness}\label{sec:swarm Cohesiveness}
The necessity of swarm dispersal which enables the agents to spread and explore is clear. However, dispersal may lead to undesired swarm separation, resulting in multiple distinct sub-swarms. This section discusses algorithms proposed by Barel et. al. \cite{barel2019come} on a variety of gathering algorithms. These algorithms by their nature prevent swarm separation.

Ando et al. \cite{ando1999distributed} showed that a pair of mutually visible agents $a_i, a_j$ with a visibility range $V_a$ will keep their visibility intact as long as they both move within a common symmetrical consensus area.
This area is a circumscribed disk $d_{ij}$, with a radius $\frac{1}{2}V_a$, centered at the agent's midpoint location $m_{ij}$. If they move according to this constraint their distance never exceeds $V_a$, and agents preserve mutual visibility.
That yields the following:

\begin {equation}
    \parallel\vec{p}_i(t) - \vec{p}_j(t)\parallel\leq V_a \Rightarrow \parallel\vec{p}_i(t+1) - \vec{p}_j(t+1)\parallel\leq V_a
    \label{eq: Distances_in_AR}
\end{equation}

\begin{figure}[!ht]
    \begin{minipage}[t]{0.37\columnwidth}
        \centering
        \fbox{\includegraphics[width=\columnwidth]{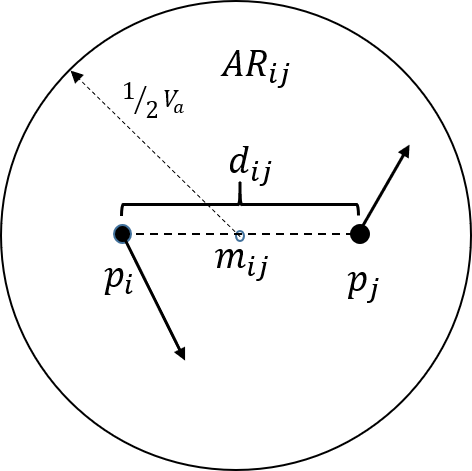}}
        \captionof{subfigure}{Two agents, $a_i,a_j$ having visibility range $V_a$, located within $d_{ij}$ disk with a radius $\frac{1}{2}V_a$, centered in their midpoint $m_{ij}$.}
        \label{fig: Allowable Region a}
    \end{minipage}\hfill
    \begin{minipage}[t]{0.53\columnwidth}
        \centering
        \fbox{\includegraphics[width=\columnwidth]{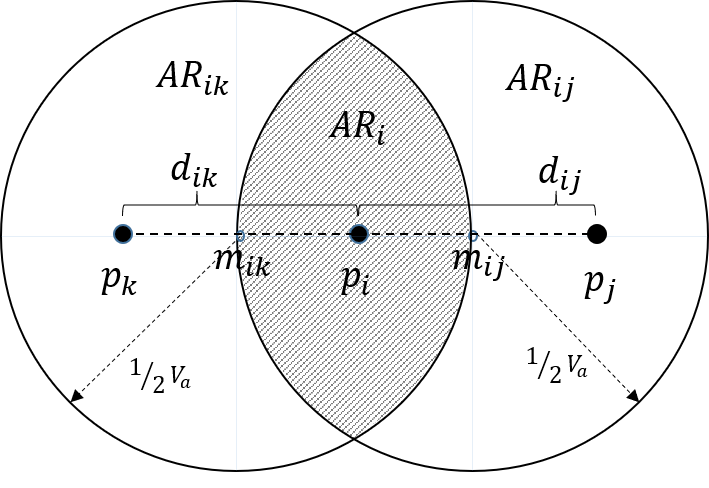}}
        \captionof{subfigure}{The agent's $a_i$ allowable region is the intersection of its allowable regions $AR_{ik}, AR_{ij}$, and is depicted by the dotted area $AR_i$.}
        \label{fig: Allowable Region b}
    \end{minipage}
    \caption{$AR_i$ Description of Allowable Region \cite{manor2019local}.}
    \label{fig:allowable_regions}
\end{figure}

Manor et al. \cite{manor2019local} defined this common area as ``Allowable Region" (AR) where connectivity is preserved:
 $$AR_{ij}\triangleq d_{ij} = d_{ji}\triangleq AR_{ji}$$
To keep on agent's connectivity to several chosen neighbors $N_i$, the allowable region, denoted $AR_i$, is the intersection of the $AR$ with those neighbors, as illustrated in by Fig. \ref{fig:allowable_regions}
\begin{equation}
    AR_i=\bigcap\limits_{j\in N_{i}}AR_{ij}
\label{eq:ARs}
\end{equation}

\subsubsection{Swarm Flexibility} \label{sec: Swarm Flexibility}
This section delves into two strategies for maintaining cohesion and their impact on swarm formation: the ``Never Lose a Neighbor" approach and a more flexible algorithm referred to as ``Flexible Swarms", which has been integrated into our model.\\ 

\textbf{Never Lose a Neighbor}\label{sec:never_lose_a_neighbor}: A familiar approach for keeping swarm cohesiveness is to apply the policy of ``never lose a neighbor", where agents that once sensed each other preserve their visibility relationship forever.
Let us define $N_i$ as a set of agents that are neighbors of agent $a_i$, i.e agents that are no more than $V_a$ from $a_i$:
 
\begin{equation}
    \begin{split}
        & j \in N_i \Leftrightarrow \parallel p_i-p_j\parallel \leq V_a \\
        & \forall i \;\: \text{and} \;\: \Delta t \geq 0 : \quad j \in N_{i}(t) \Rightarrow j \in N_{i}(t+\Delta t)\\
    \end{split}
    \label{eq: Never Lose a Friend}
\end{equation}
This policy is quite conservative and imposes significant constraints on the agents' mobility, reducing the sensorial coverage area of the swarm during exploration.\\

\textbf{Flexible Swarms:} To relax the ``Never Lose a Neighbor" policy we incorporated the ``Flexible Swarms" algorithm by Manor et al. \cite{manor2019local}, into the formation layer. This algorithm ensures that the swarm is kept connected, but not necessarily fully interconnected, by allowing some of the connections in the visibility graph to be trimmed, without endangering the graph’s connectivity\footnote{Graphs are often used to describe interactions, such as visibility between agents in multi-agent systems. Graphs are commonly labeled as $G(\mathcal{V},\mathcal{E})$, as $\mathcal{V}$ = $\{\nu_{1}, \nu_{2},..., \nu_{n}\}$ is the set of vertices representing the agents, and $\mathcal{E} \subseteq \mathcal{V} \times \mathcal{V}$ is the set of edges representing mutual visibility between agents. The neighborhood set of a vertex $v_{i}$ is the set of vertices connected to it, i.e. $N_{i} \triangleq \{\nu_{j}\in \mathcal{V}\;$ $|\;$ $\{\nu_{i},\nu_j \} \in \mathcal{E} \}$.}. 

This algorithm is based on Toussaint’s work on Relative Neighborhood Graphs (RNG) \cite{toussaint1980relative} 
 which reduces the graph's number of edges while maintaining connectivity.

The algorithm output is a set of certain neighbors, noted as “Effective Neighbors”, within the agent's local environment, and with whom the agent needs to keep connectivity thus ensuring that the swarm stays connected.

Let us denote the set of effective neighbors with whom agent $a_i$ must keep connection, as $N^{e}_{i}.$\\
\begin{equation}
\forall j \in N_{i} \\\ : \\\ j \in N^{e}_{i} \iff \nexists k : \left\{\begin{alignedat}{3}
    & \|p_{i}-p_{k}\| < \|p_{i}-p_{j}\| \\
    & \& \\
    & \|p_{j}-p_{k}\| < \|p_{i}-p_{j}\|\\
  \end{alignedat}\right.
\label{eq:EffectiveNeighbor}
\end{equation}

\subsubsection{Attraction-Repulsion Interaction}\label{sec: Attraction Repulsion Interaction}
 This algorithm motivates agents to spread out, seek, and manage tasks over the plane as a result of embedded repulsion forces.
 Swarm scattering is essential for exploration. This feature is achieved by an agent's behavior (found in the agent layer) when it interacts with its neighbors. An example of similar behaviour can be found in \cite{del2018importance}. Such distributed and decentralized task allocation interaction functions are described in \cite{amir2023multi} where exploration is derived from inter-agent repulsion and task attraction influence functions.
The local dynamics rule based on the influence functions is: 
\begin{equation}
\begin{split}
    \vec{p}_i(t\hspace{-0.1cm}+\hspace{-0.1cm}1) \hspace{-0.1cm}& = \vec{p}_i(t) - \delta \frac{\vec{v}_i}{\lVert\vec{v}_i\lVert} \\
    \vec{v}_i & = \hspace{-0.1cm}\sum_{\substack{j = 1}}^{\mathcal{N}} \hspace{-0.08 cm}F_r(\lVert \vec{p}_i \text{-} \vec{p}_j \rVert)\overrightarrow{p_i p_j} -\hspace{-0.15cm}\sum_{\substack{k = 1}}^{\mathcal{K}}\hspace{-0.08 cm} F_a(\lVert \vec{p}_i \text{-} \vec{c}_k \rVert) \overrightarrow{p_i c_k}\\
\end{split}
\label{eq:attractionrepulsiondynamicsgoalgeneral}
\end{equation}
where $\delta$ represents the maximal step size of the agents and $F_r$,$F_a$ are the repulsion and attraction functions respectively.
Although \cite{amir2023multi} presents a well-simulated distributed task allocation behaviour, a constraint of this system is that it requires the agents to operate within a bounded area and the swarm to have a number of agents proportional to that area. In this paper, we overcome these limitations and avoid them.
\subsubsection{Peristaltic Mechanism}\label{sec:The Peristaltic Effect}
\begin{figure}[ht]
    \begin{center}
    \begin{minipage}[t]{0.37\columnwidth}
        \fbox{
            \includegraphics[width=\columnwidth]{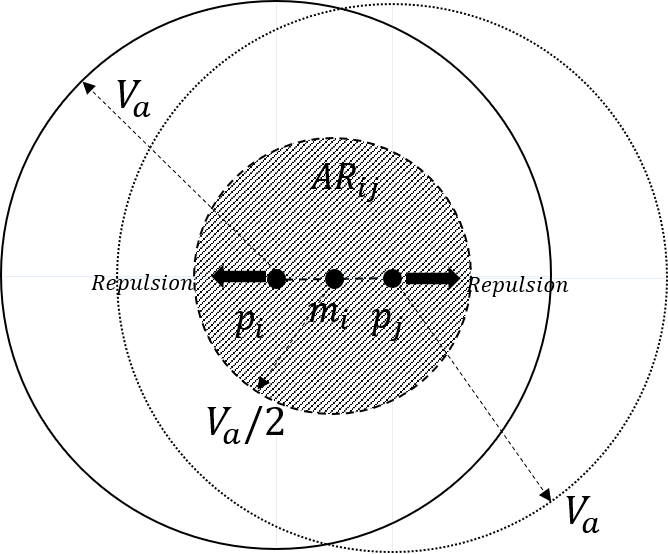}
        }    
    \end{minipage} \hspace{0.4 cm}
    \begin{minipage}[t]{0.455\columnwidth}
        \fbox{
            \includegraphics[width=\columnwidth]{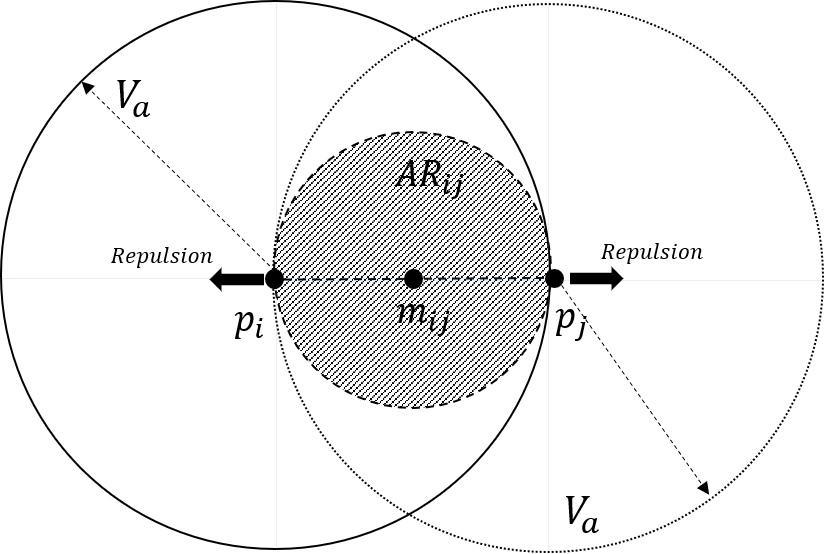}
        }
    \end{minipage} \vspace{0.5mm} % Add vertical space between rows       
\caption{ 
Two agents $a_i,a_j$ repel each other when mutually visible. The left image depicts the agents' movement direction within $AR_{i,j}$ as a result of this repulsion.
The right image illustrates the agents' locations after some time, where they are positioned on the boundary of their $AR$ and can maneuver only on their $AR's$ circumference.}
\label{fig:MmovingOnCircumference}
\end{center}
\end{figure}

As the swarm spreads, the repulsion interactions between agents motivate the agents to move away from each other within their AR.
Over time, as the agents drift away one from another, they reach positions where they can move only along the perimeter of their respective AR, as illustrated in Fig. \ref{fig:MmovingOnCircumference}. As a result, the swarm's maneuvering capability deteriorates, potentially leading to fixation or very limited movements.

\begin{figure}[!ht]
    \centering
    \begin{minipage}[t]{0.30\columnwidth}
        \centering
        \includegraphics[width=\columnwidth]{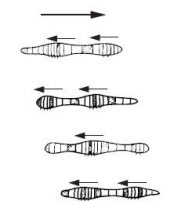}
        \captionsetup{labelformat=empty}
        \captionof{subfigure}\centering{Worm}
        %\label{fig: worm}
    \end{minipage}
    \begin{minipage}[t]{0.35\columnwidth}
        \centering
        \includegraphics[width=\columnwidth]{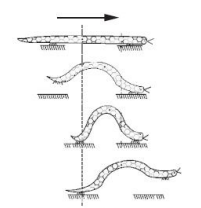}
        \captionsetup{labelformat=empty}
        \captionof{subfigure}\centering{Caterpillar}
        %\label{fig: caterpillar}
    \end{minipage}
    \begin{minipage}[t]{0.30\columnwidth}
        \centering
        \includegraphics[width=\columnwidth]{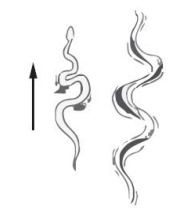}
        \captionsetup{labelformat=empty} % Apply the new label format
        \captionof{subfigure}\centering{Snake}
        %\label{fig: snake}
    \end{minipage}
    \caption{Bio-mimic of locomotion techniques of limbless animals:  worm, caterpillar and snake (source: \cite{juhasz2013analysis}).}
    \label{fig: Locomotion techniques of limbless animals}
\end{figure}

A physiological process called the Peristalsis inspired us to incorporate a bio-mimic algorithm that operates in a similar manner. The physiological process involves alternating contractions and relaxations of the smooth muscles creating a wave-like motion that enables forward movement. An example of peristalsis can be observed in worm movement \cite{juhasz2013analysis} as illustrated in Fig.\ref{fig: Locomotion techniques of limbless animals}, where creating relaxation at the rear part of the body enabling the front part to move forward and vice versa.

Our algorithm similarly enables some relaxation of the influence of the repulsion interactions. 
We implement peristaltic movement by adding a small random step within $AR$ as illustrated in Fig.\ref{fig:Peristaltic and  random walk}.
This stochastic behavior is also beneficial in preventing deadlocks and overcoming local minima, where the interaction influences zero out and movement motivation is formed. %The random walk is also known as a starting point for the many analytical studies of animal dispersal movement \cite{okubo1986dynamical}.
\vspace{1cm}
\begin{figure}[!ht]
%\captionsetup{width=85mm}
  \centering
    \includegraphics[width=5cm]{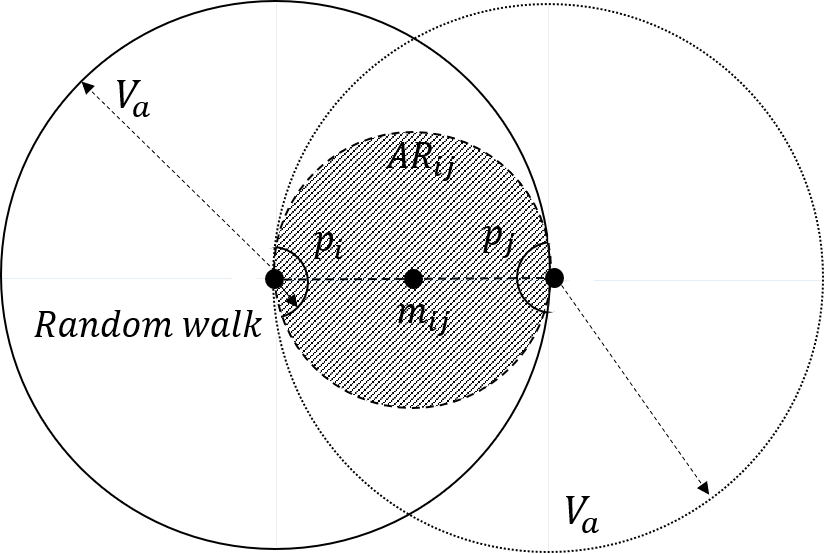}
    \caption{Adding random walk steps for simulating the peristaltic effect within the agents' $AR$.}
    \label{fig:Peristaltic and random walk}
\end{figure}
\vspace{-0.5 cm}
\subsection{The Motion Rules}\label{sec: The Motion Rules}
By integrating these algorithms, we introduce a local motion rule that ensures swarm connectivity even in an infinite plane, without preset boundaries. This enhancement is necessary to mimic the behavior of animal swarms in the wild, where often the area's boundaries are not predetermined externally.

Our multi-layered swarm behavior model yields the following sequence of individual local motion rules:
%Ariel Table formation ***

\begin{center}
\line(1,0){230}\\
Algorithms 1: Local motion Rule\\
\vspace{-0.25 cm}
\line(1,0){230}\\
\begin{enumerate}
%\item For each agent $a_i$ at time $t$ do
%    \item \begin{equation} Find effective neighbors $N_i^e$ according to Eq. (\ref{eq:EffectiveNeighbor}) \label{Algo} \end{equation}
    \vspace{-0.25cm }\item For each agent $a_i$ at time $t$ do
    \vspace{-0.25cm }\item Find effective neighbors $N_i^e$ according to Eq. \hspace{-0.2 cm} (\ref{eq:EffectiveNeighbor}) 
    \vspace{-0.25cm }\item Calculate $AR_i$ with $N_i^e$, according to Eq. (\ref{eq:ARs})
    \vspace{-0.25cm }\item Calculate $\vec{p}_i(t+1)$ by Eq. (\ref{eq:proposedperistalticattractionrepulsiondynamics}) and Eq. (\ref{eq:nextmove}) 
    \vspace{-0.25cm }\item End
    \vspace{-0.5 cm}
\end{enumerate}
\line(1,0){230}
\end{center}
where,
\begin {equation}
\begin{split}
    %\vec{p}_i(t+1) & = \vec{p}_i(t) - \delta  \frac{\vec{v}_i}{\lVert\vec{v}_i\lVert} \\
    \vec{p}_{i_{proposed}} (t) & = \vec{p}_i(t) - \delta  \frac{\vec{v}_i}{\lVert\vec{v}_i\lVert} \\
    \vec{v}_i & = \vec{V}_{i_{repulsion}} + \vec{V}_{i_{attraction}} \\
    \vec{V}_{i_{repulsion}} & = \sum_{\substack{j = 1}}^{\mathcal{N}} F_r(\lVert \vec{p}_i - \vec{p}_j\rVert)\overrightarrow{p_i p_j} \\ 
    \vec{V}_{i_{attraction}} & = -\sum_{\substack{k = 1}}^{K} F_a(\lVert \vec{p}_i - \vec{c}_k \rVert) \overrightarrow{p_i c_k} \\
\end{split}
\label{eq:proposedperistalticattractionrepulsiondynamics}
\end{equation}

Given that the agent needs to be confined and to maneuver exclusively within $AR_i$, if, at any time, the agent's forthcoming position $\vec{p}_{proposed}$ extends beyond its $AR_i$ the agent's next position is the projected $\vec{p}_{proposed}$ on its $AR_i$ boundary according to $\overrightarrow {p_i(t)p_{proposed}}$.
\begin{equation}
\vec{p}_i(t+1)=\begin{cases}
\vec{P}_{i_{proposed}} + \vec{r} & \hspace{-2.3cm} \text{if }\vec{P}_{proposed}\in AR_i \vspace {0.4 cm}\\

\overrightarrow{p_i(t)p_{i_{proposed}}} \cap \text{perimeter}(AR_i) + \vec{r}& \hspace{-0.2cm}\text{o.w.}
\end{cases}
\label{eq:nextmove}
\end{equation}
where $\delta$ is the agent's step size in each time step and $\vec{r}$ is the peristaltic step within $AR_i$.\\
\section{Broadcast Control of the Swarm}\label{sec:Steering and Guiding the Swarm}
As the swarm seeks for tasks in its environment, these tasks might be positioned at a considerable distance, extending past its current sensory field. This situation may occur as well after fulfilling the tasks' requirements (i.e. assuming the tasks disappeared) and continue exploring for new ones. While a random walk might eventually allow the swarm to detect these tasks, directing the swarm toward the desired goal area would be more efficient.

Steering and controlling a swarm of agents through centralized control poses a significant challenge. Synchronizing with each agent via communication requires real-time wide-band and long-range communication and central computing resources that are not scalable, insufficiently robust, and may be exposed to a single point of failure.
It is generally agreed upon that most animals in the wild operate with decentralized independence, and not under any central control.

An alternative approach is a distributed and decentralized steering control, where agents with limited sensory range operate autonomously and may be guided by an external observer as introduced by Barel et al. \cite{barel2018steering}: An external observer guides a swarm of identical and indistinguishable agents, despite the agents lacking information on global location and orientation. This is achieved through simple broadcast signals, based on the observed swarm average location, limiting the need for sending specific information to individual agents. 

In this paper, we present a guiding concept we call \textbf{``Anonymous Leaders"}. This concept is inspired by the biological behavior of shoals, as described by Reebs \cite{reebs2000can}, which shows that the foraging movements of a shoal derive from the leadership of a few knowledgeable individuals, eliciting following behavior rather than the collective will of the shoal mates. 
 
The concept of ``Anonymous Leaders" defines a sub-group of agents that are informed about the direction of tasks located beyond their sensory capabilities. These agents are attracted to the remote tasks while keeping cohesion with other swarm members. The agents gain their information either by enhanced sensory capabilities or guidance signals received from an external observer as in \cite{barel2018steering}. By this, we relax our initial paradigms by letting all agents have the ability to receive only broadcast signals with no explicit inter-agent communication.
The leaders are anonymous, interacting with the adjacent agents indistinguishably. The leaders obey the local motion rules described in section (\ref{sec: The Motion Rules}) while taking into consideration the attraction influence of the goal area in Eq.(\ref{eq:proposedperistalticattractionrepulsiondynamics}) as shown in Eq.(\ref{eq:leaderproposedperistalticattractionrepulsiondynamics}).
\begin {equation}
\begin{split}
    %\vec{p}_i(t+1) & = \vec{p}_i(t) - \delta \frac{\vec{v}_i}{\lVert\vec{v}_i\lVert} \\
    \vec{p}_{i_{proposed}} (t) & = \vec{p}_i(t) - \delta \frac{\vec{v}_i}{\lVert\vec{v}_i\lVert} \\
    \vec{v}_i & = \vec{V}_{i_{attraction}} \\ 
    \vec{V}_{i_{attraction}} & = - F(\lVert \vec{p}_i - \vec{p}_{\text{goal area}} \rVert) \overrightarrow{p_i p_{\text{goal area}}} \\
\end{split}
\label{eq:leaderproposedperistalticattractionrepulsiondynamics}
\end{equation}

As these leaders aim to step toward the goal area, the rest of the swarm, as a result of cohesion conservation, follows. 

\section {Parameterization and simulations}
By simulation, we conducted tests and analyses on various swarm configurations, each exhibiting distinct behaviors. These configurations were achieved by modifying the following input key parameters:
\begin{itemize}
    \item Swarm size: This parameter determines the number of agents constituting the swarm.
    \item Tasks constellation and requirements: These parameters establish the positions of tasks and their respective demands. This information is initially unknown to the agents and needs to be explored.
    \item Flexibility policy (Sec \ref{sec: Swarm Flexibility}): This binary parameter determines the swarm's neighbor connectivity policy, offering the choice between ``Never Lose a Neighbor" or the ``Flexible Swarm" policy for each run. 
    \item Inter-agent interaction (Sec \ref{sec: Attraction Repulsion Interaction}): This binary parameter defines whether the repulsion interaction function between agents is activated upon mutual visibility or not.
    \item Magnitude of the Peristaltic Mechanism (Sec \ref{sec:The Peristaltic Effect}) - This parameter sets the random step size.
    \item Tasks area positioning for guidance (Sec \ref{sec:Steering and Guiding the Swarm}): This parameter dictates the location of the tasks, which is solely known to the observer, if exists, responsible for guiding the swarm.
\end{itemize}

The simulation's output key parameters are:
\begin{itemize}
    \item Swarm position and swarm exploration capability: These parameters are calculated and represented by the center of the swarm's bounding circle and the radius of the bounding circle itself.
    \item Distance to task area: This output parameter is based on the distance between the swarm position and the designated task.
    \item Target satisfaction: This output parameter describes how many tasks are currently fulfilled.    
\end{itemize}

\section {Results and discussion}
In this section, we examine and demonstrate the contribution of each of the algorithms to the swarm's behaviour as well as point out key-parameters tradeoffs.
\subsection {Swarm exploration and flexibility}
\begin{figure}[!ht]
    \centering
    \begin{minipage}[t]{0.30\columnwidth}
        \centering
        \includegraphics[width=\columnwidth]{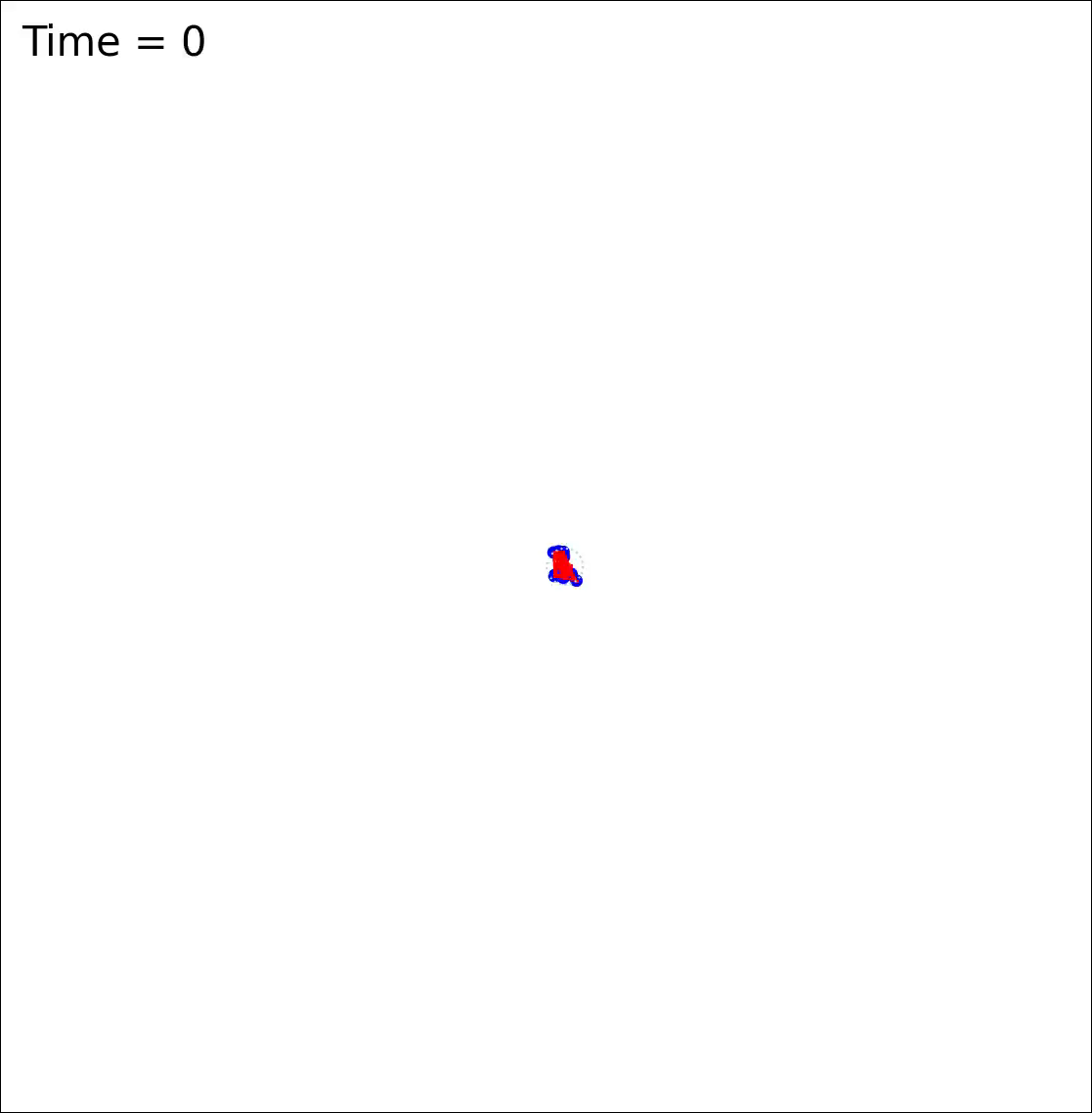}
        %\captionof{subfigure}\centering{}
    \end{minipage} \vspace{0.5mm} % Add vertical space between rows
    \begin{minipage}[t]{0.30\columnwidth}
        \centering
        \includegraphics[width=\columnwidth]{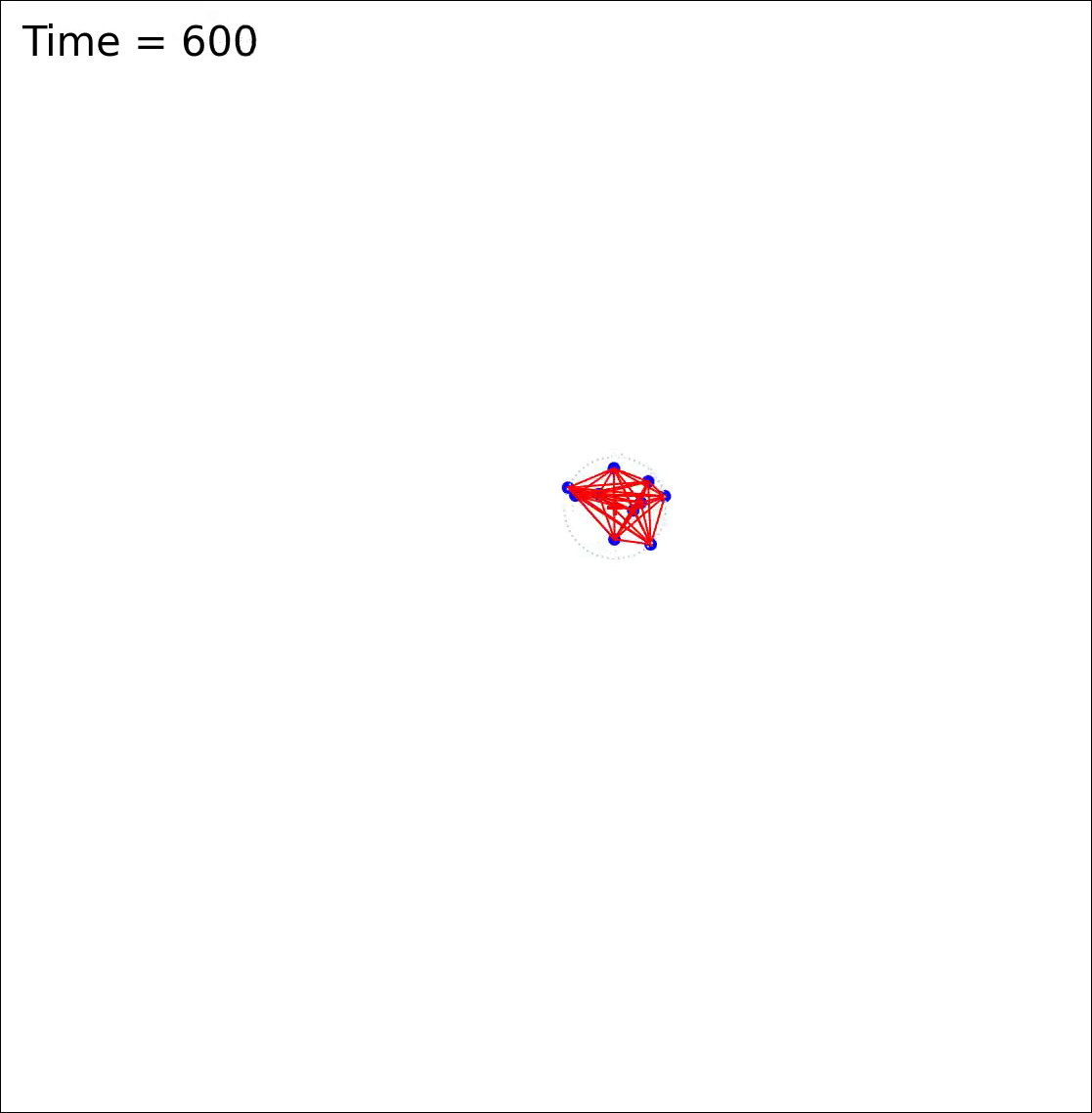}
        %\captionof{subfigure}\centering{}
    \end{minipage}
    \begin{minipage}[t]{0.30\columnwidth}
        \centering
        \includegraphics[width=\columnwidth]{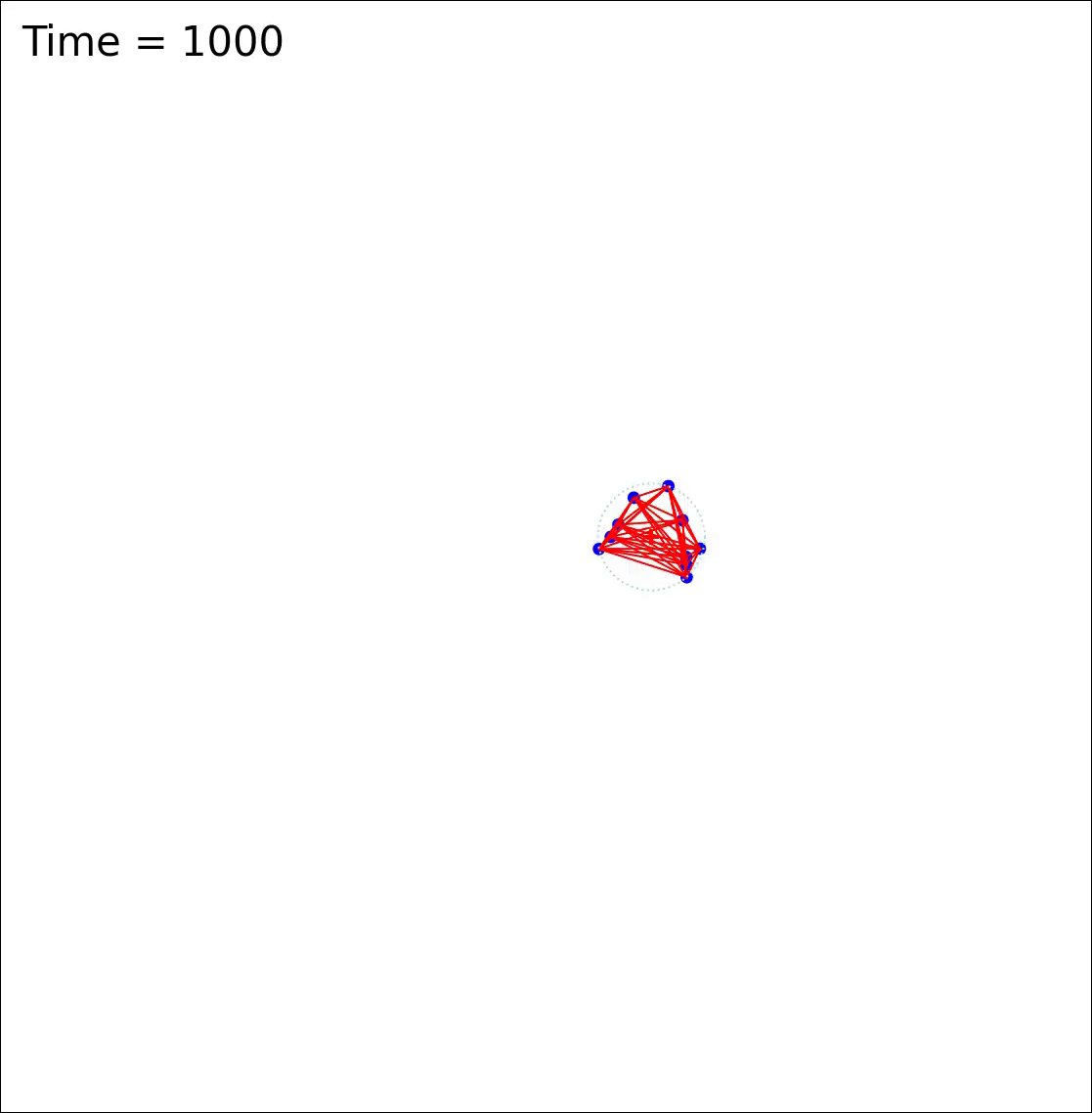}
        %\captionof{subfigure}\centering{}
    \end{minipage}
    \begin{minipage}[t]{0.30\columnwidth}
        \centering
        \includegraphics[width=\columnwidth]{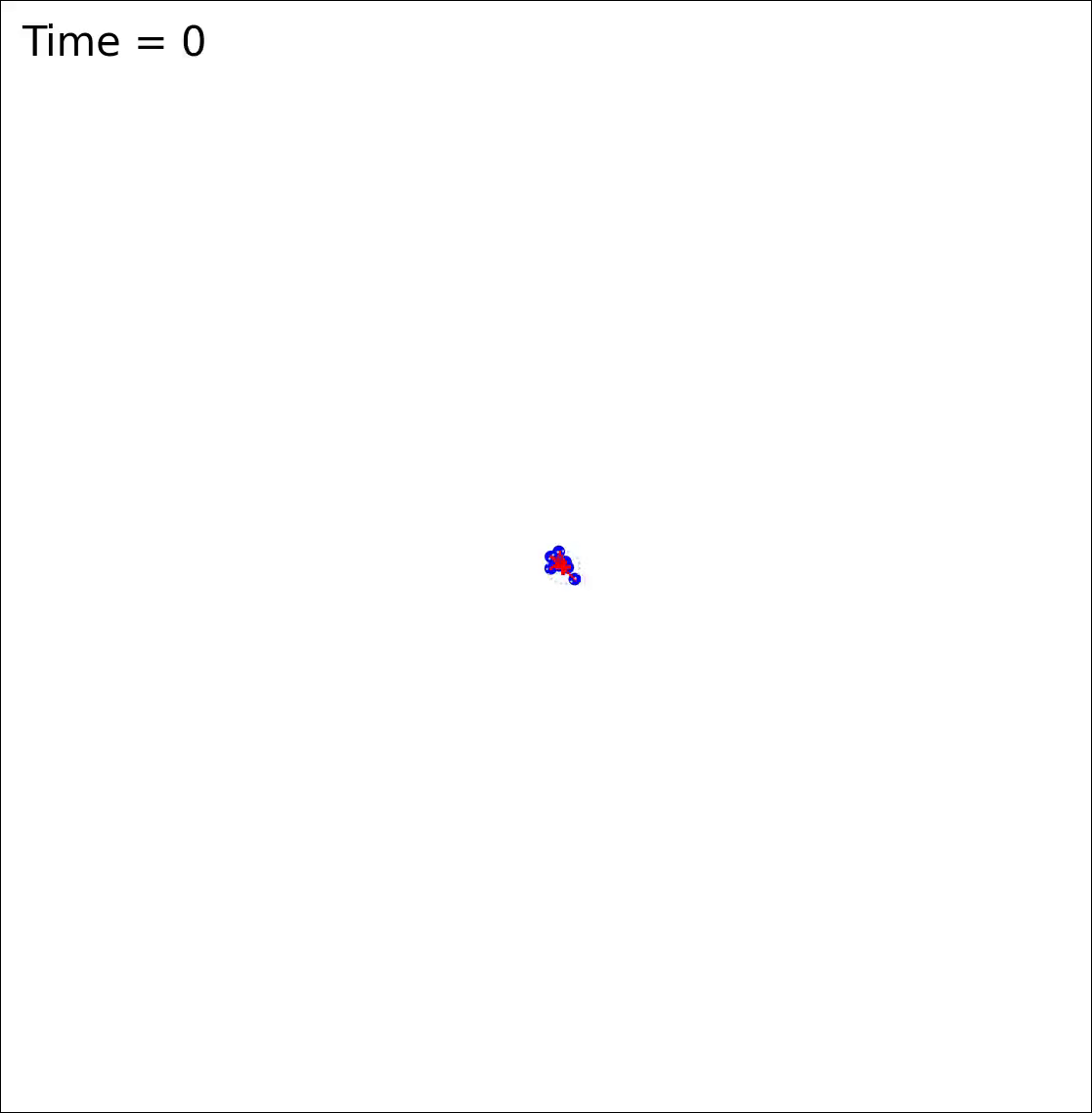}
        %\captionof{subfigure}\centering{}
    \end{minipage}
    \begin{minipage}[t]{0.30\columnwidth}
        \centering
        \includegraphics[width=\columnwidth]{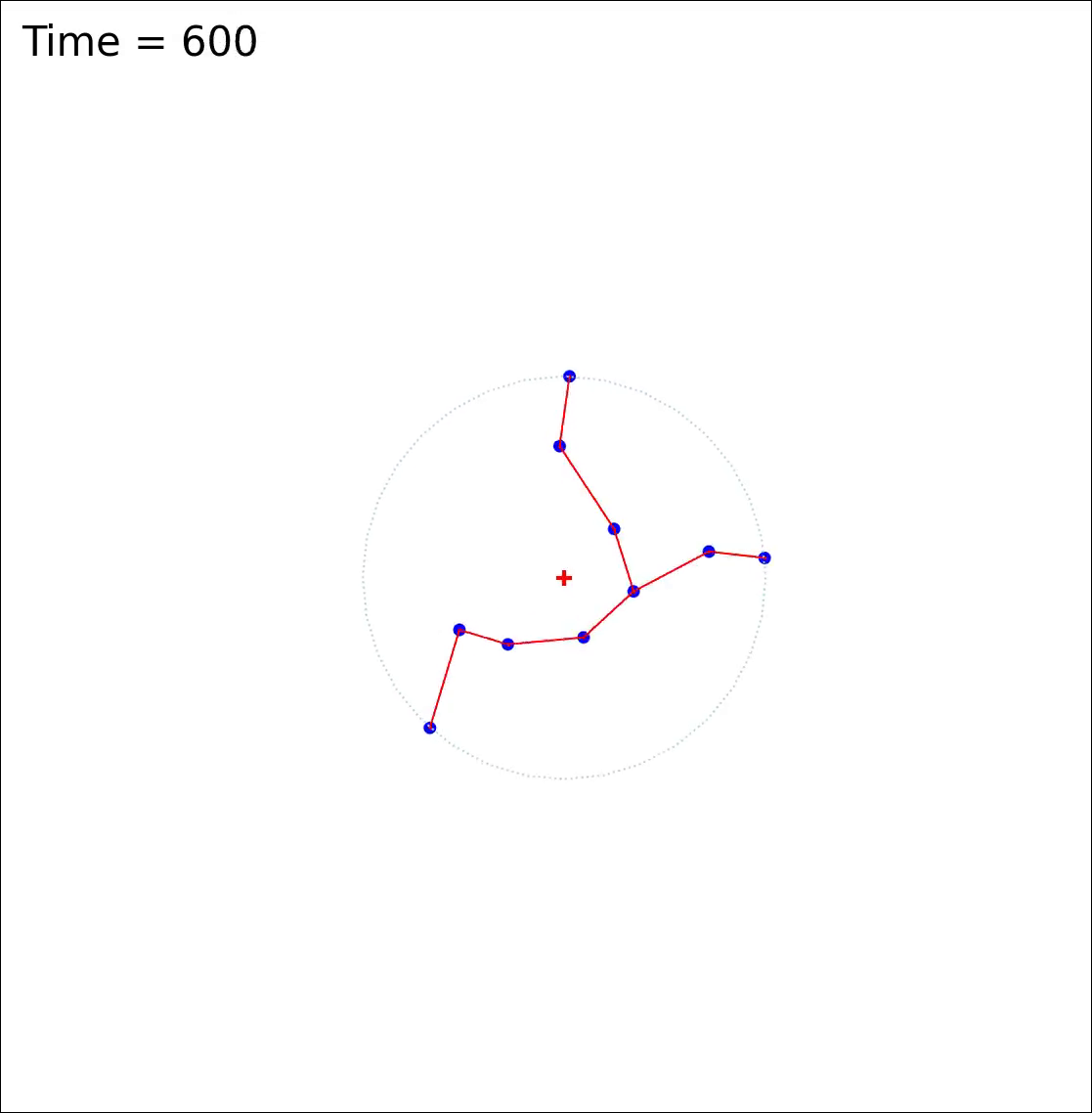}
        %\captionof{subfigure}\centering{}
    \end{minipage}
    \begin{minipage}[t]{0.30\columnwidth}
        \centering
        \includegraphics[width=\columnwidth]{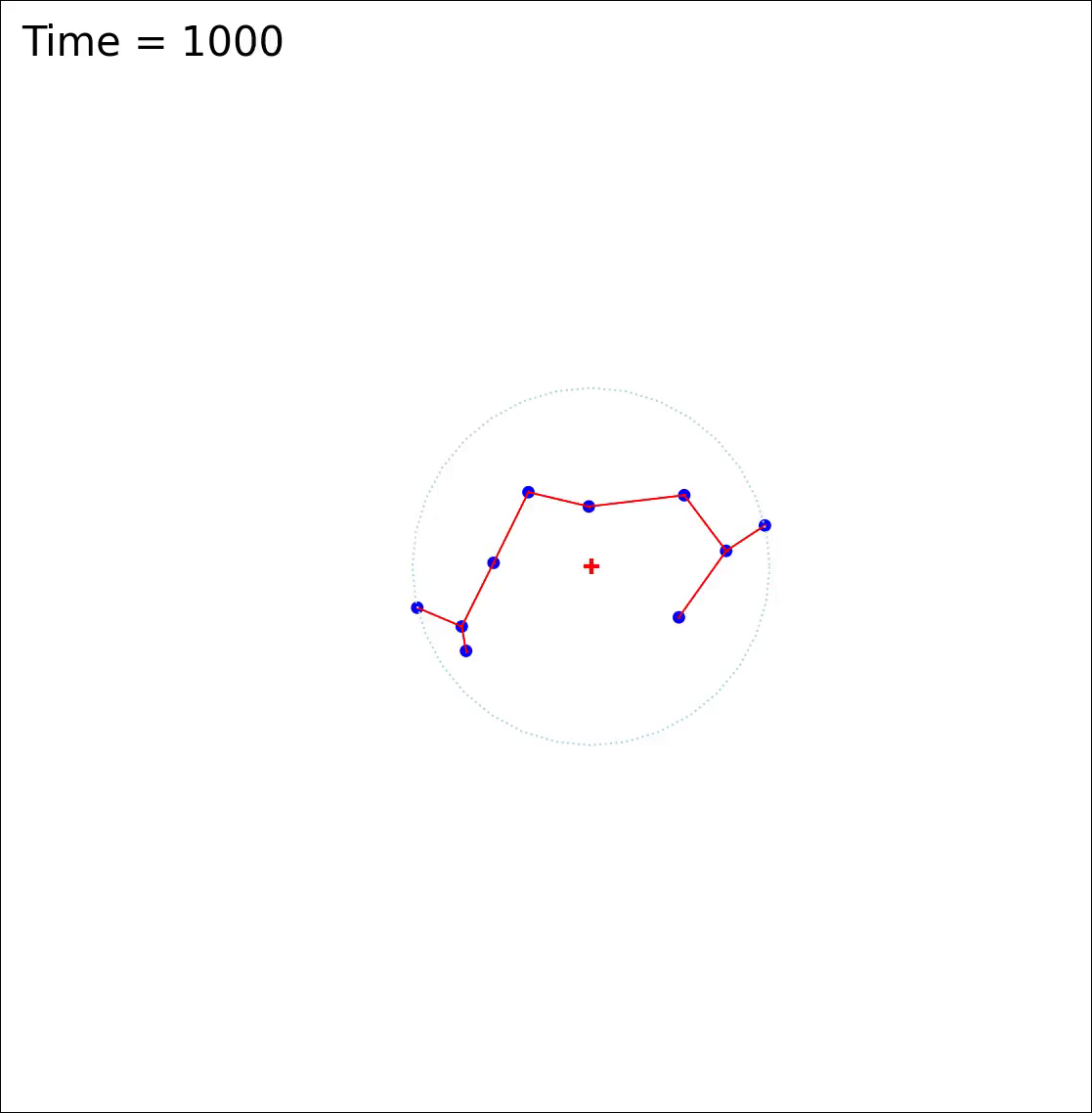}
        %\captionof{subfigure}\centering{}
        %\label{fig: snake}
    \end{minipage}\vspace{0.8mm}
    \begin{minipage}[t]{0.95\columnwidth}
        \centering
        \includegraphics[width=\columnwidth]{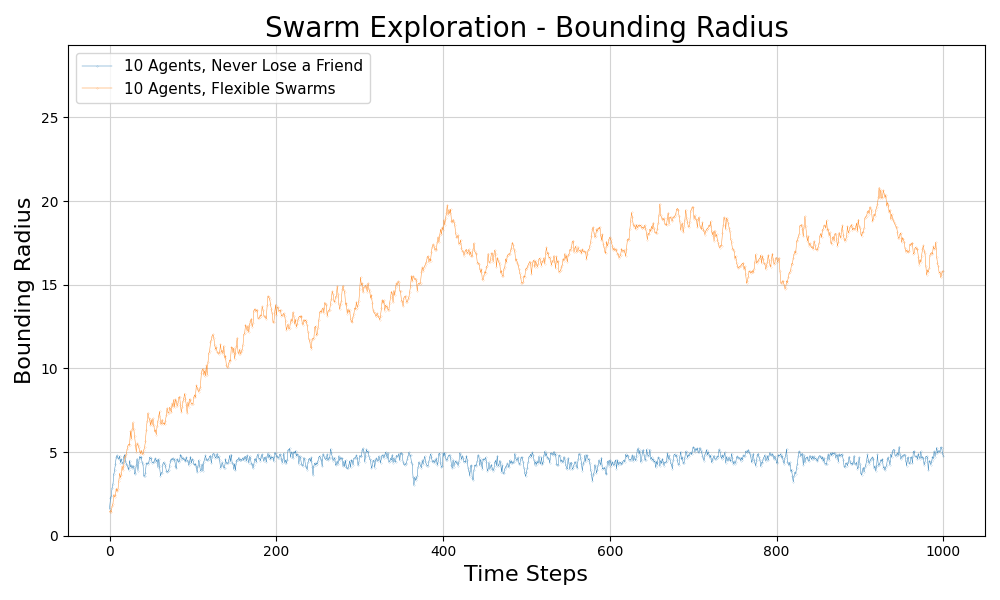}
        %\captionof{subfigure}\centering{}
        %\label{fig: snake}
    \end{minipage}
    \caption{
    The influence of flexibility on the swarm size over time (from left to right). The figure shows two rows of images: the top row displays swarm exploration under the "Never Lose a Friend" algorithm, while the bottom row illustrates exploration under the "Flexible Swarms" algorithm. The comparative bounding radius graph illustrates the difference between the two algorithms throughout the entire scenario.}
    \label{fig:Nlif and flexible swarm}
\end{figure}

In this scenario, we examined the impact of the swarm flexibility policy on its exploration capability. We conducted simulation runs of a swarm operating under two distinct behaviors: ``Never Lose a Neighbor" and ``Flexible Swarms" (Sec \ref{sec: Swarm Flexibility}), as depicted in Fig. \ref{fig:Nlif and flexible swarm}. The graph illustrates that the swarm's bounding radius is naturally confined to the visibility range under the ``Never Lose a Neighbor" policy. Conversely, under the ``Flexible Swarm" policy, this radius extends much farther, significantly enhancing the swarm's neighborhood exploration capabilities.

\subsection {Swarm exploration and inter-agent interactions}
\begin{figure}[!ht]
    \centering
    \begin{minipage}[t]{0.30\columnwidth}
        \centering
        \includegraphics[width=\columnwidth]{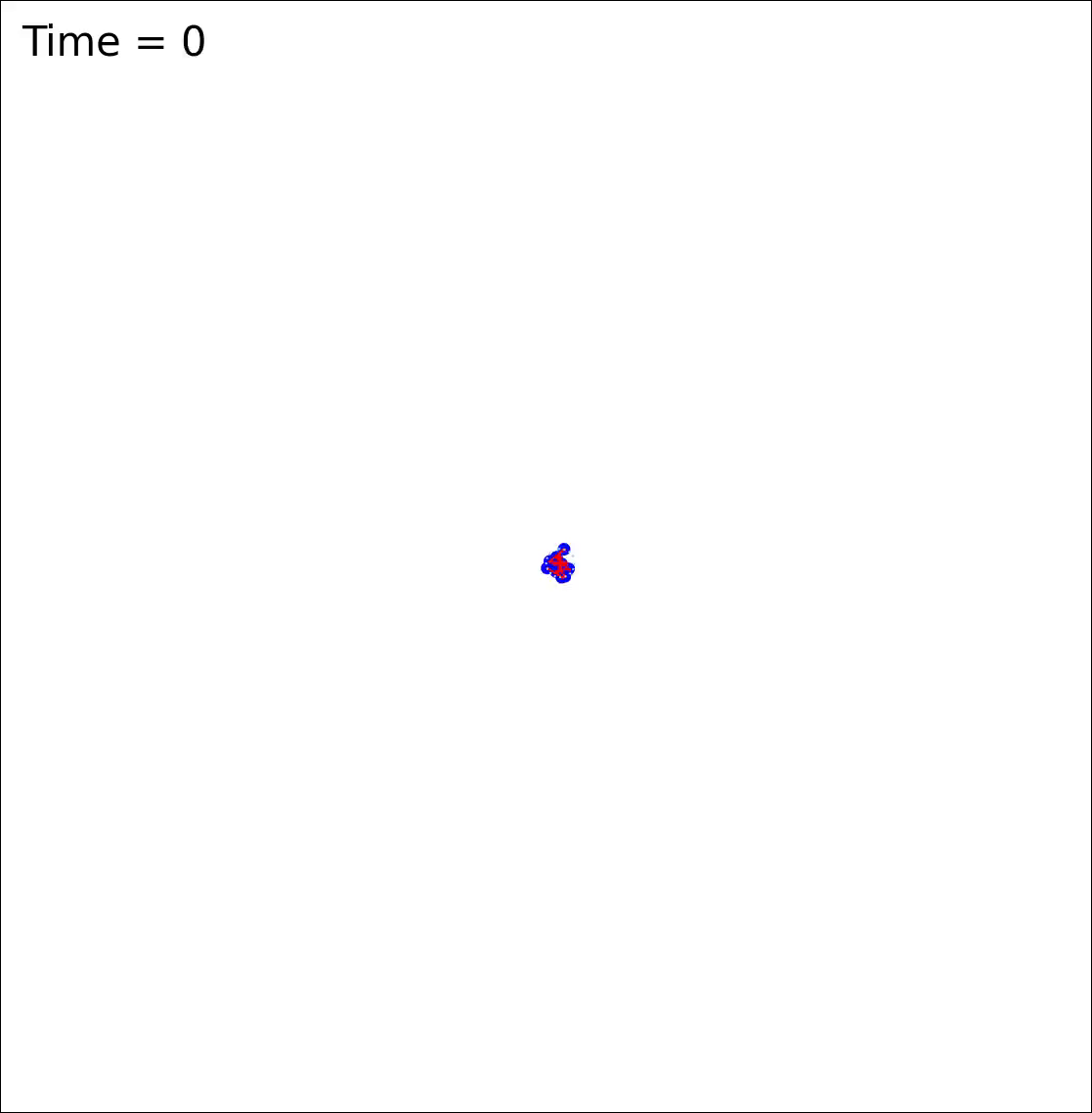}
        %\captionof{subfigure}\centering{}
        %\label{fig: }
    \end{minipage}\vspace{0.5mm}
    \begin{minipage}[t]{0.30\columnwidth}
        \centering
        \includegraphics[width=\columnwidth]{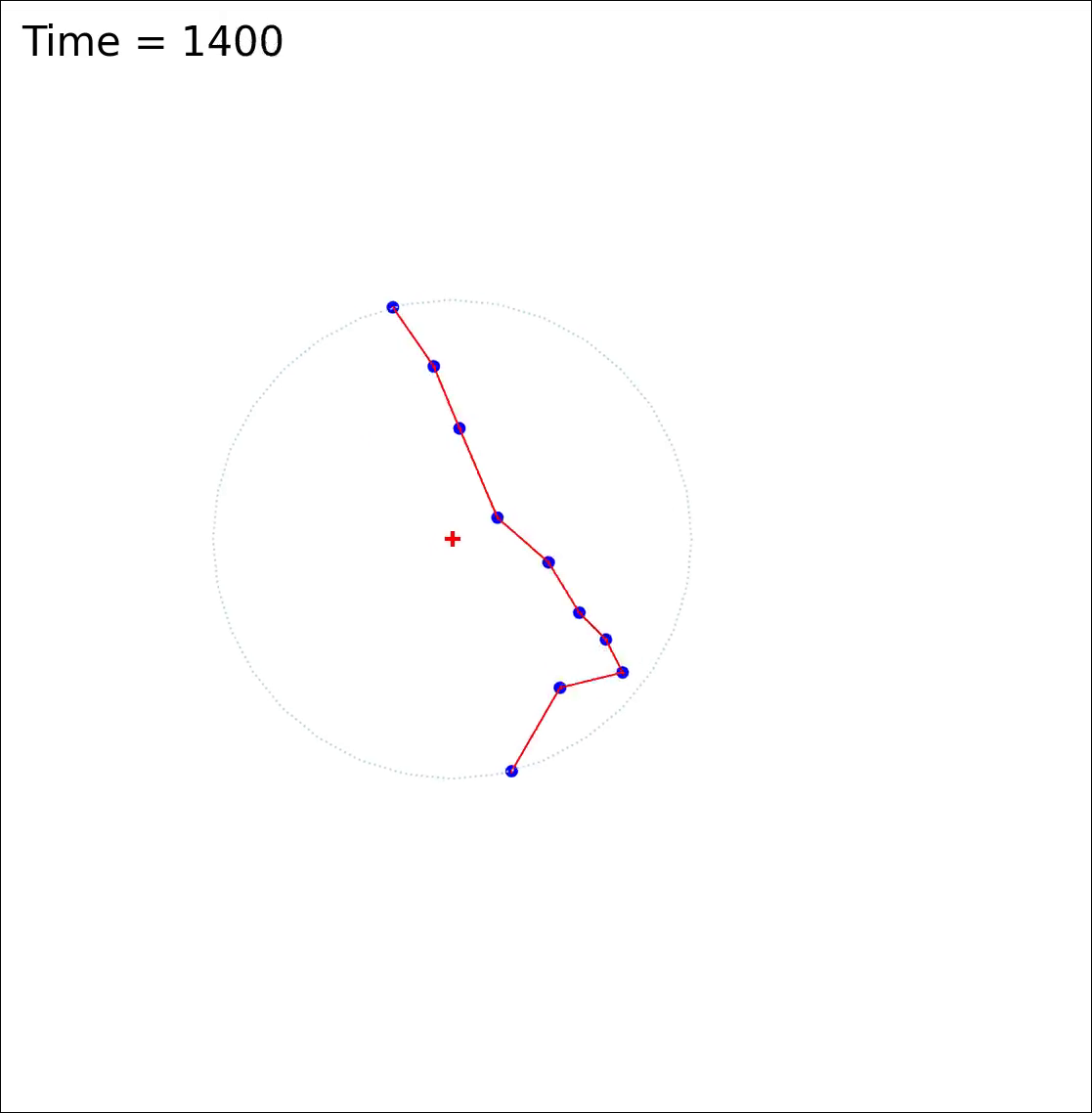}
        %\captionof{subfigure}\centering{}
        %\label{fig: caterpillar}
    \end{minipage}
    \begin{minipage}[t]{0.30\columnwidth}
        \centering
        \includegraphics[width=\columnwidth]{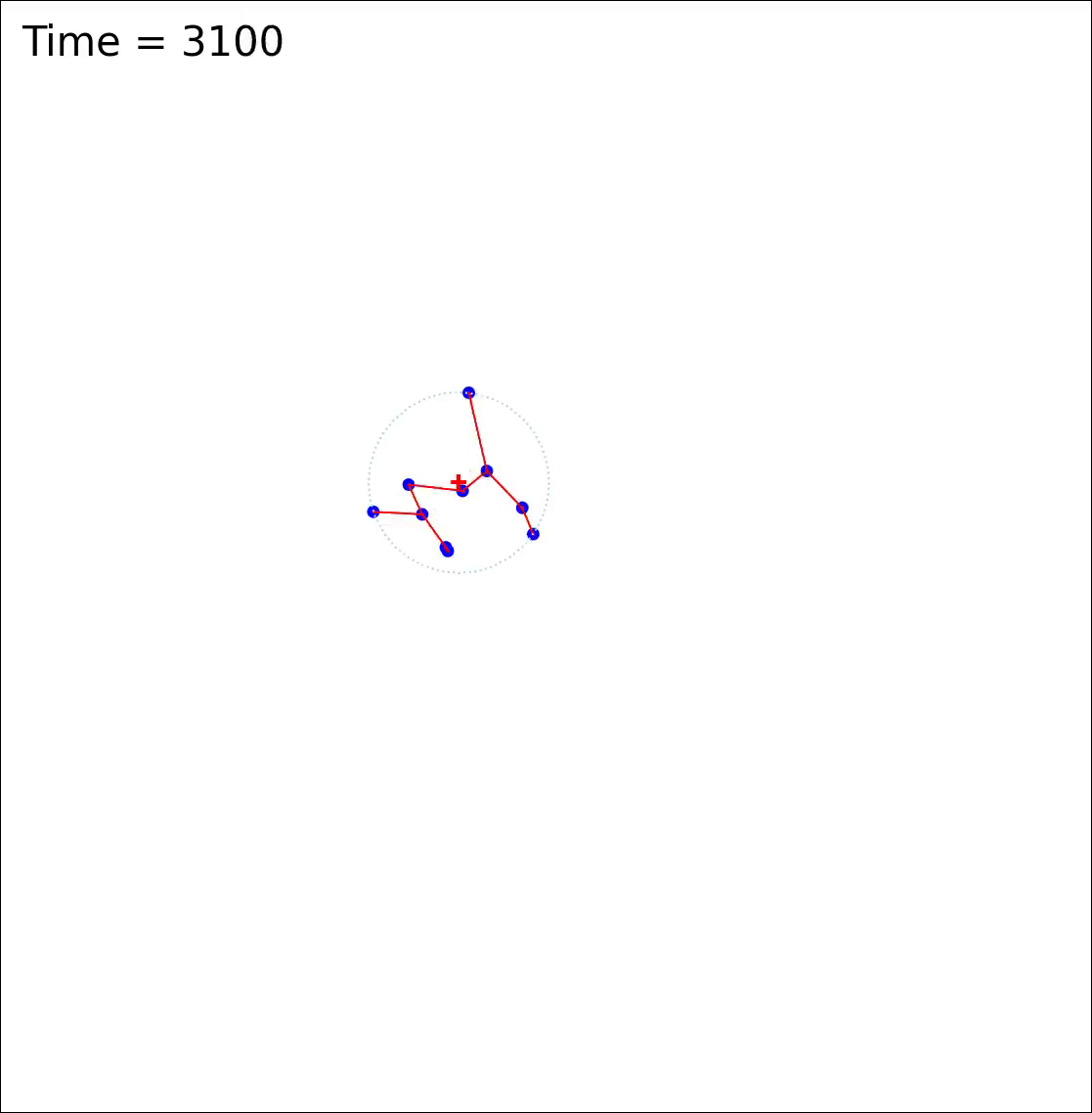}
        %\captionof{subfigure}\centering{}
        %\label{fig: snake}
    \end{minipage}
    \begin{minipage}[t]{0.30\columnwidth}
        \centering
        \includegraphics[width=\columnwidth]{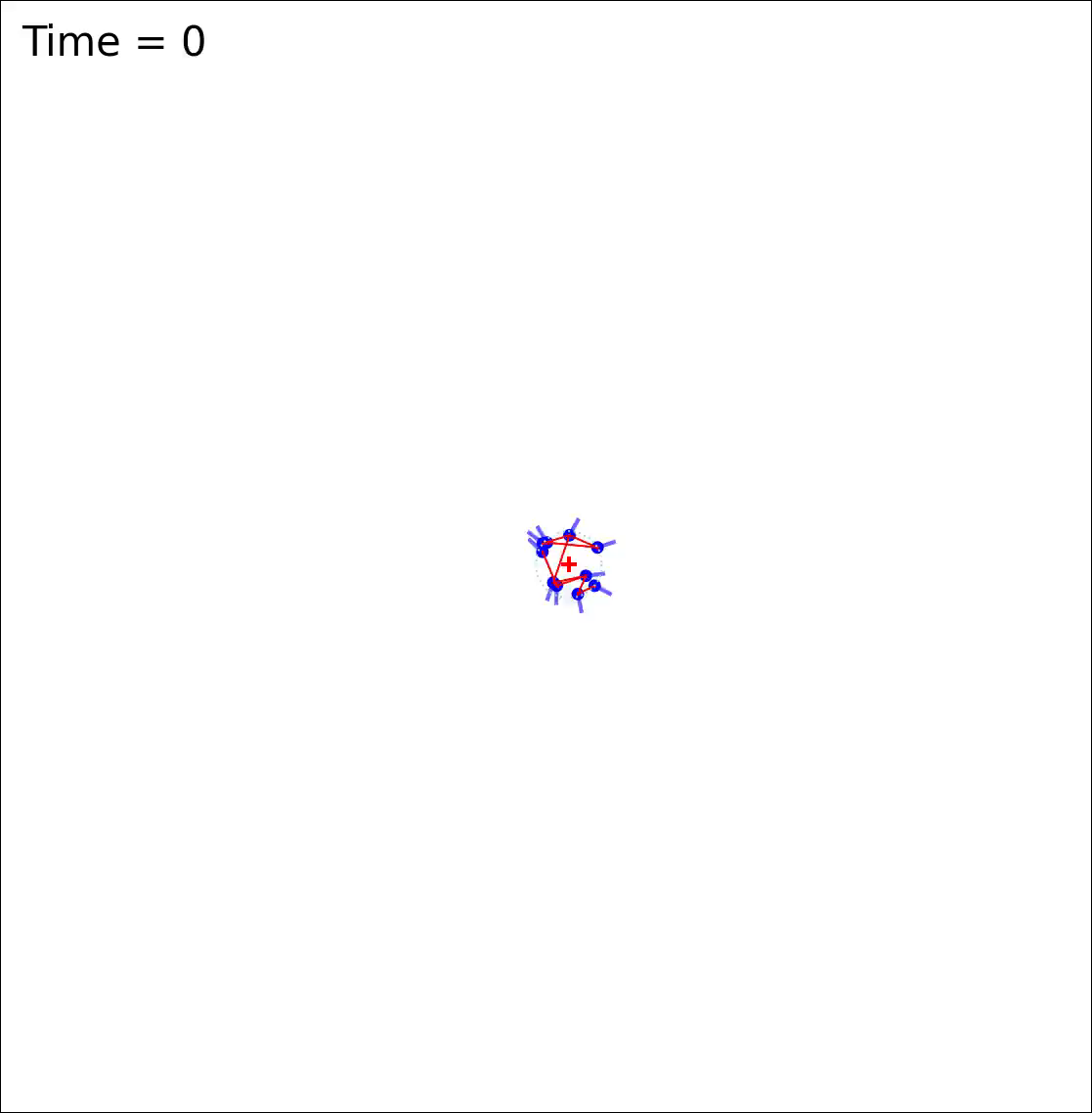}
        %\captionof{subfigure}\centering{}
        %\label{fig: snake}
    \end{minipage}
    \begin{minipage}[t]{0.30\columnwidth}
        \centering
        \includegraphics[width=\columnwidth]{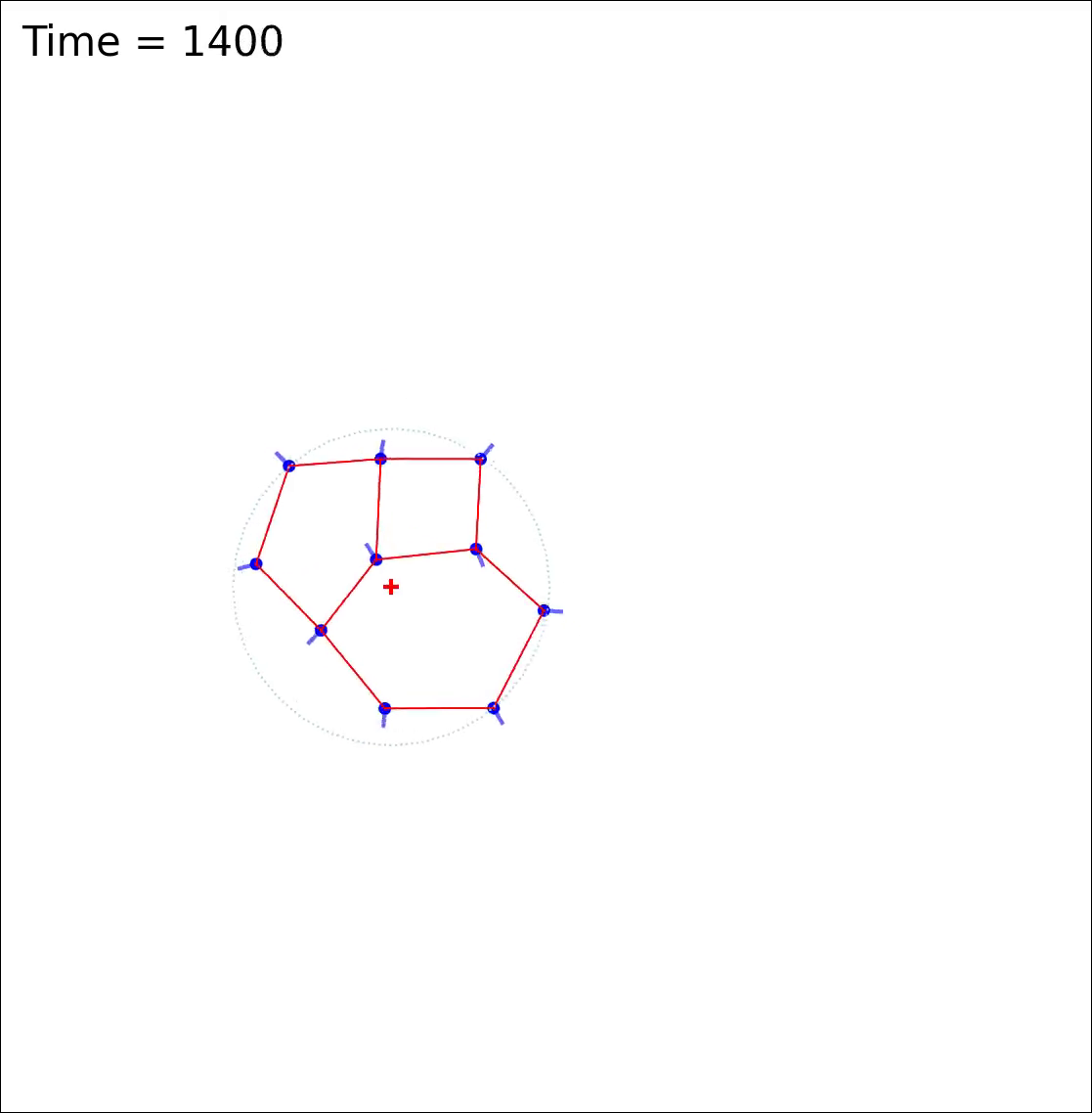}
        %\captionof{subfigure}\centering{}
        %\label{fig: snake}
    \end{minipage}
    \begin{minipage}[t]{0.30\columnwidth}
        \centering
        \includegraphics[width=\columnwidth]{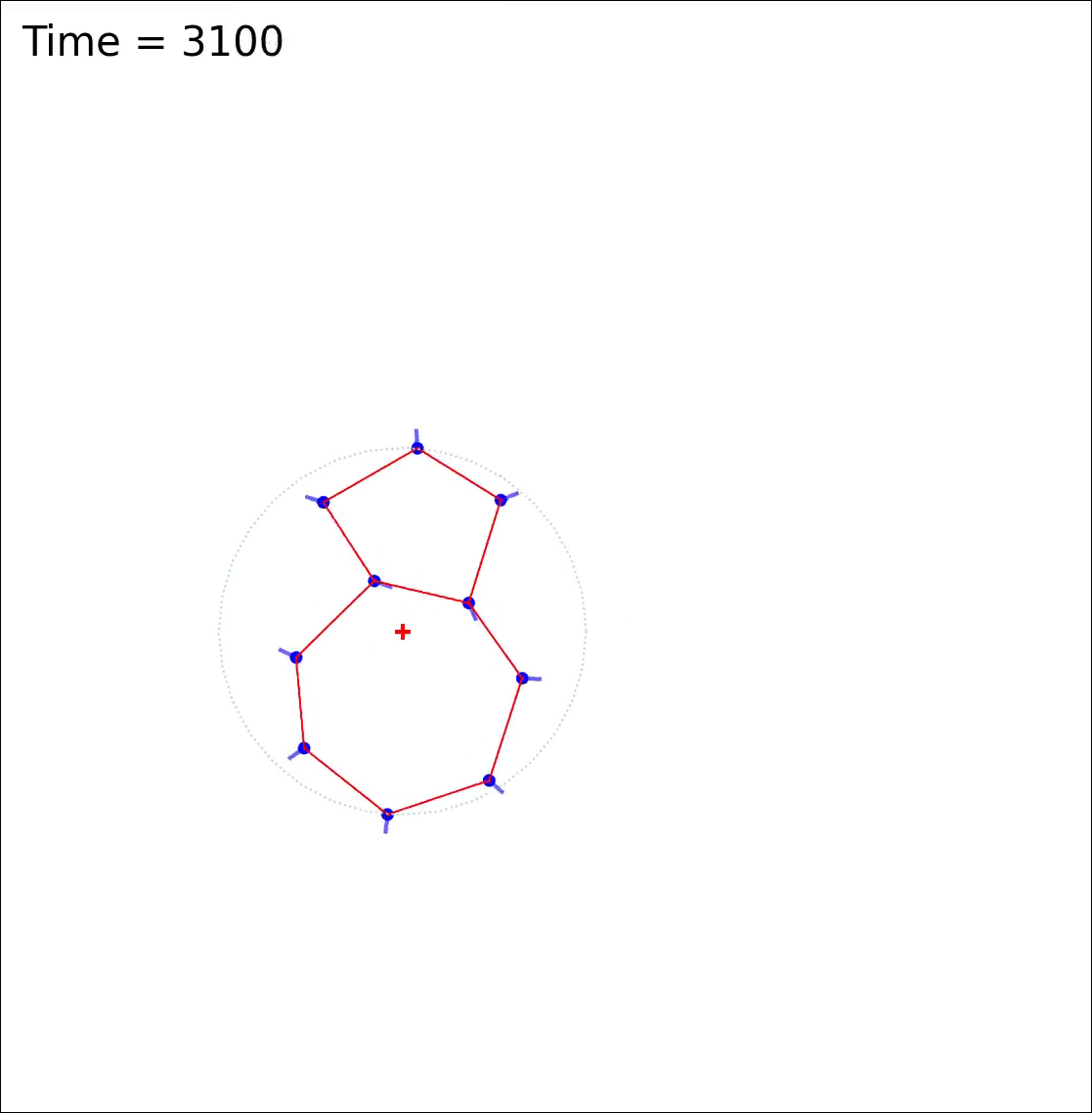}
        %\captionof{subfigure}\centering{}
        %\label{fig: snake}
    \end{minipage}\vspace{0.8mm}
    \begin{minipage}[t]{0.95\columnwidth}
        \centering
        \includegraphics[width=\columnwidth]{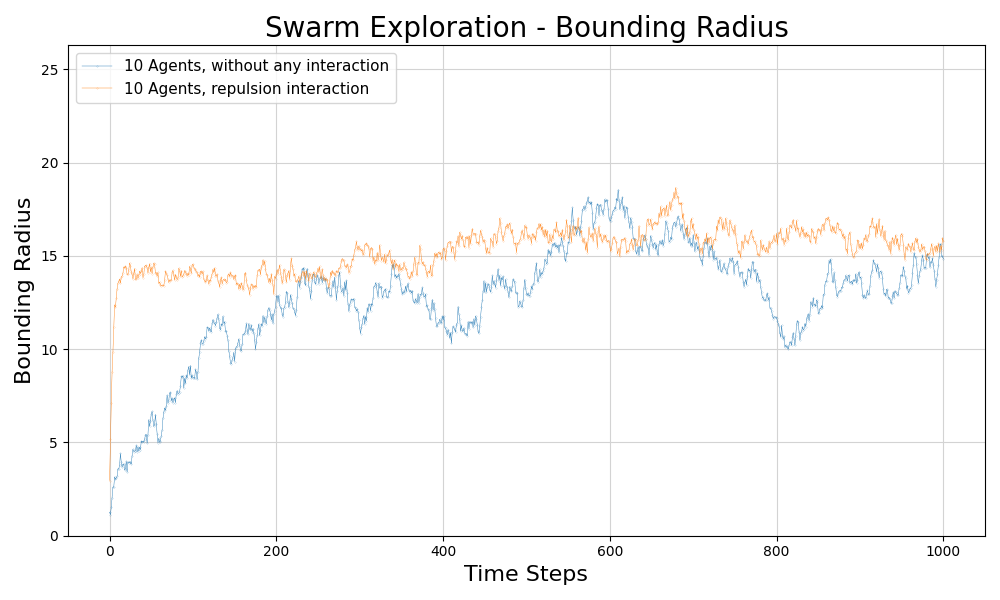}
        %\captionof{subfigure}\centering{}
        %\label{fig: snake}
    \end{minipage}
    \caption{The influence of interaction on the swarm’s behavior over time (from left to right). It consists of two rows of images for different interactions and a comparative graph between them. The top row depicts the area explored by the swarm without any repulsion interaction rule, while the bottom row depicts the exploration area under the influence of repulsion interaction. Additionally, the graph illustrates the evolution of the swarm’s bounding radius over time for each behavior.}
    \label{fig:exploration and interaction flexible swarm}
\end{figure}

In this scenario, we examined the influence of inter-agent interaction (Sec \ref{sec: Attraction Repulsion Interaction}) on the swarm exploration capabilities, assuming that attraction-repulsion interaction facilitates better environment exploration by the swarm. The results are depicted in Fig. \ref{fig:exploration and interaction flexible swarm} describing simulation runs of a swarm with and without inter-agent repulsion function. 

It's noteworthy that the repulsion interaction behavior initially induces enhanced expansion. However, over time, this effect diminishes. The graph that depicts the bounding radius of the scenario without inter-agent repulsion interaction, is considerably noisier, the bounding radius can shrink and potentially hinder exploration capability. 

\subsection {Peristaltic Mechanism}
In this section, we demonstrate the influence of the peristaltic mechanism on the maneuverability capabilities of the swarm, and how this affects the swarm's velocity under broadcast control. To generate motion, we conducted scenarios where the position of the task area is known to a specific agent, thus motivating it to move toward it. Due to the swarm's cohesiveness, the swarm generates a trial toward the goal area without explicit inter-agent communication.

\begin{figure}[!ht]
    \centering
    \begin{minipage}[t]{0.30\columnwidth}
        \centering
        \includegraphics[width=\columnwidth]{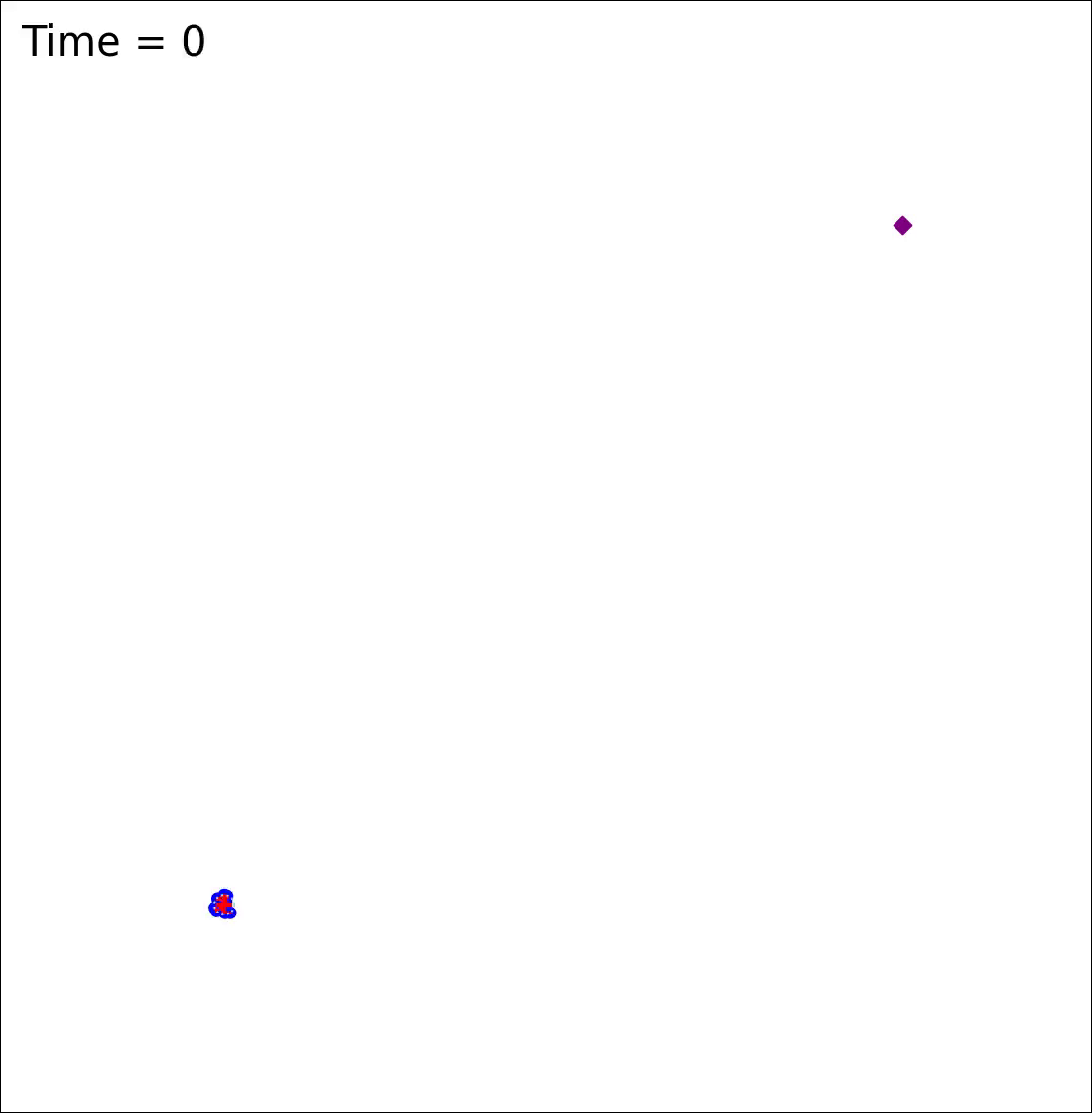}
        %\captionof{subfigure}\centering{}
    \end{minipage}\vspace{0.5mm}
    \begin{minipage}[t]{0.30\columnwidth}
        \centering
        \includegraphics[width=\columnwidth]{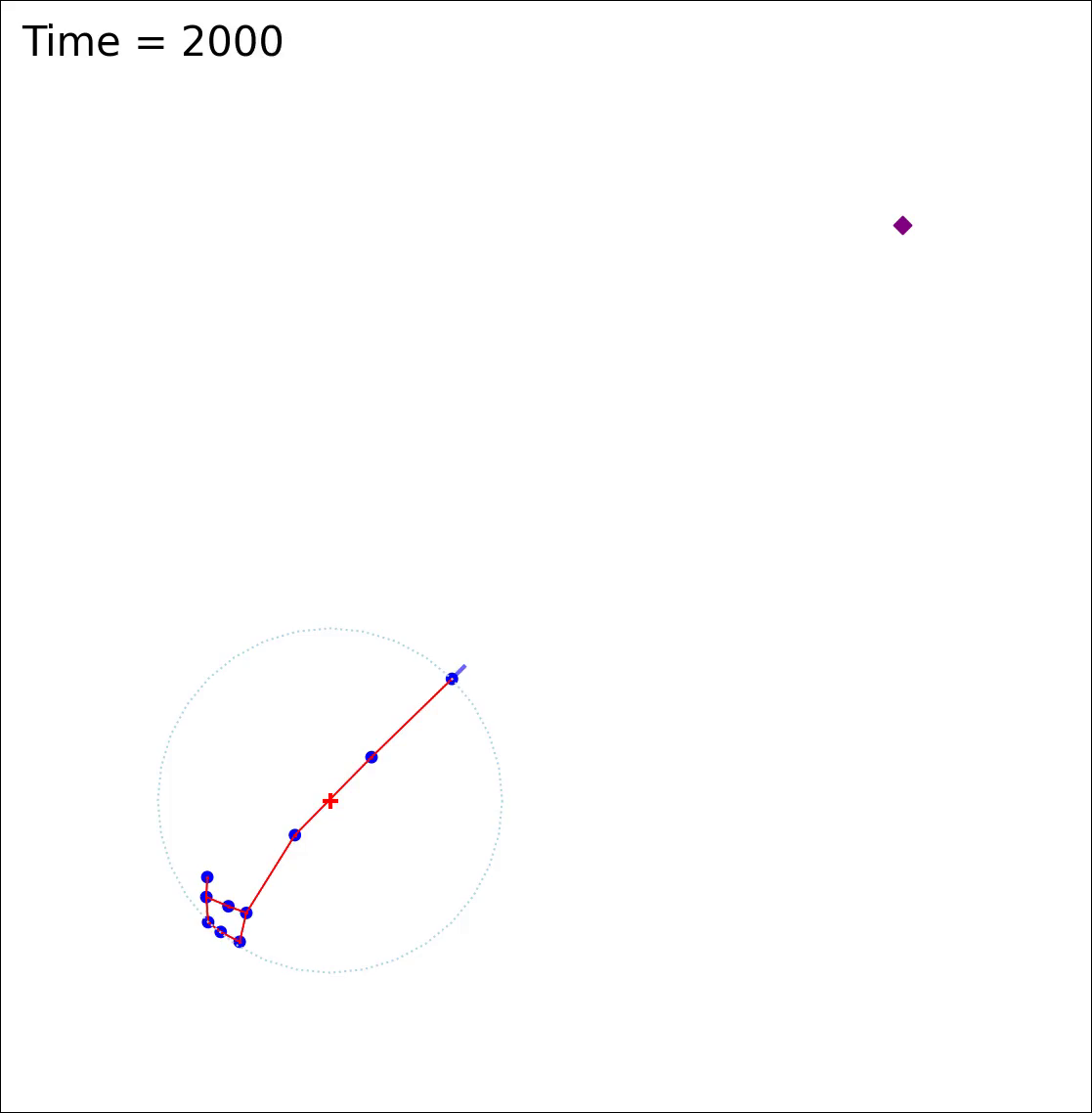}
        %\captionof{subfigure}\centering{}
    \end{minipage}
    \begin{minipage}[t]{0.30\columnwidth}
        \centering
        \includegraphics[width=\columnwidth]{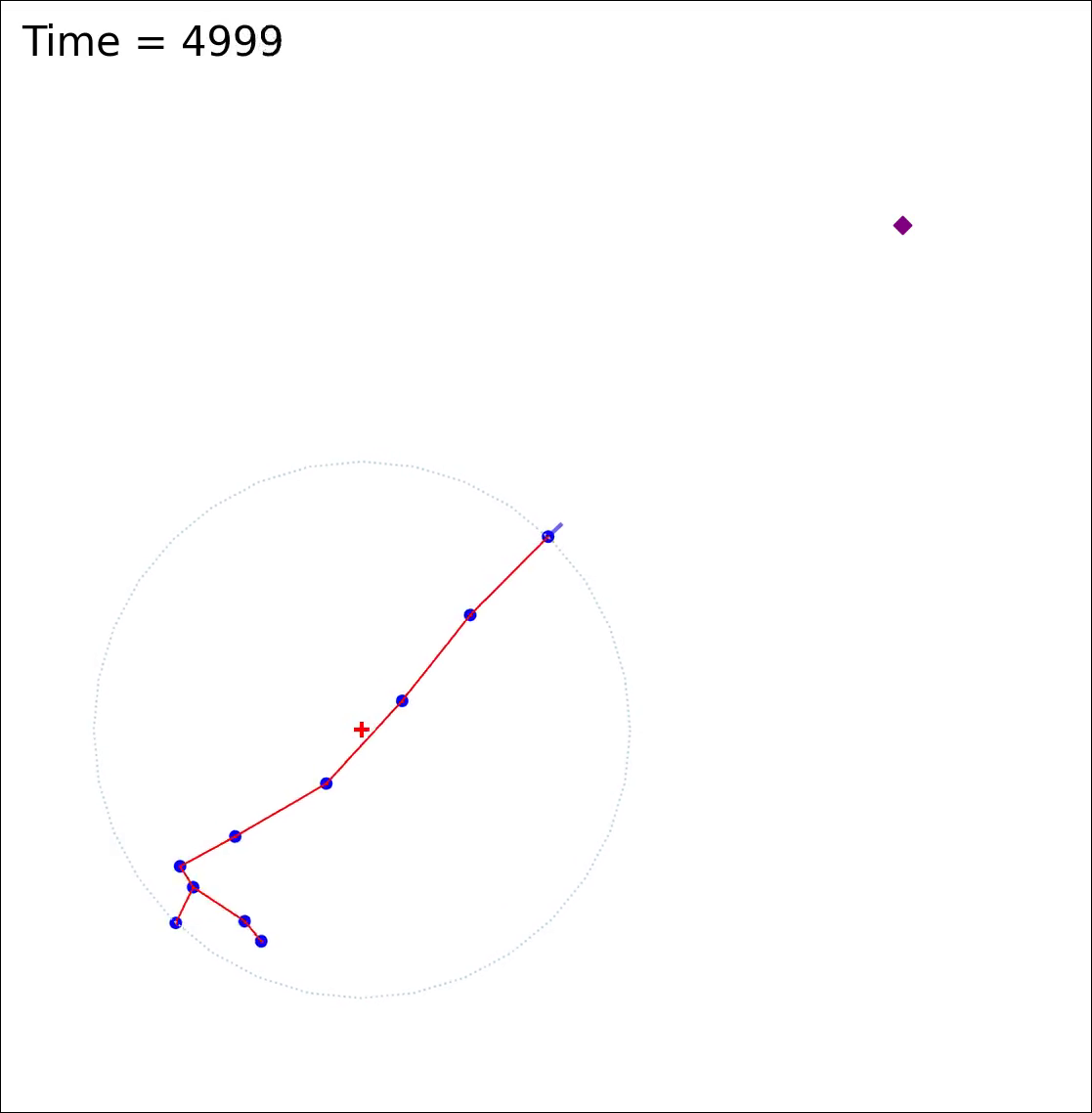}
        %\captionof{subfigure}\centering{}
    \end{minipage}
    \begin{minipage}[t]{0.30\columnwidth}
        \centering
        \includegraphics[width=\columnwidth]{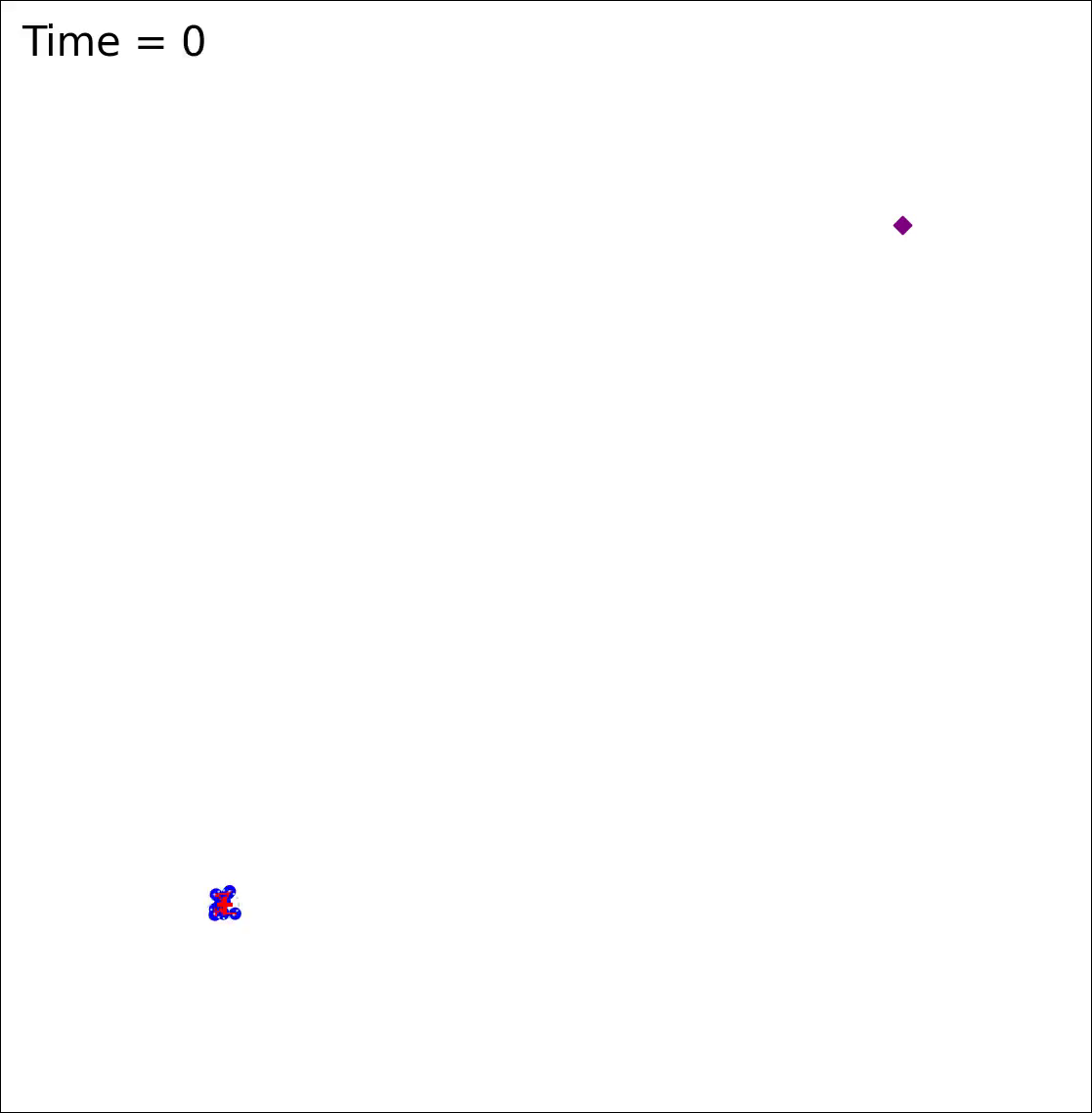}
        %\captionof{subfigure}\centering{}
    \end{minipage}\vspace{0.5mm}
    \begin{minipage}[t]{0.30\columnwidth}
        \centering
        \includegraphics[width=\columnwidth]{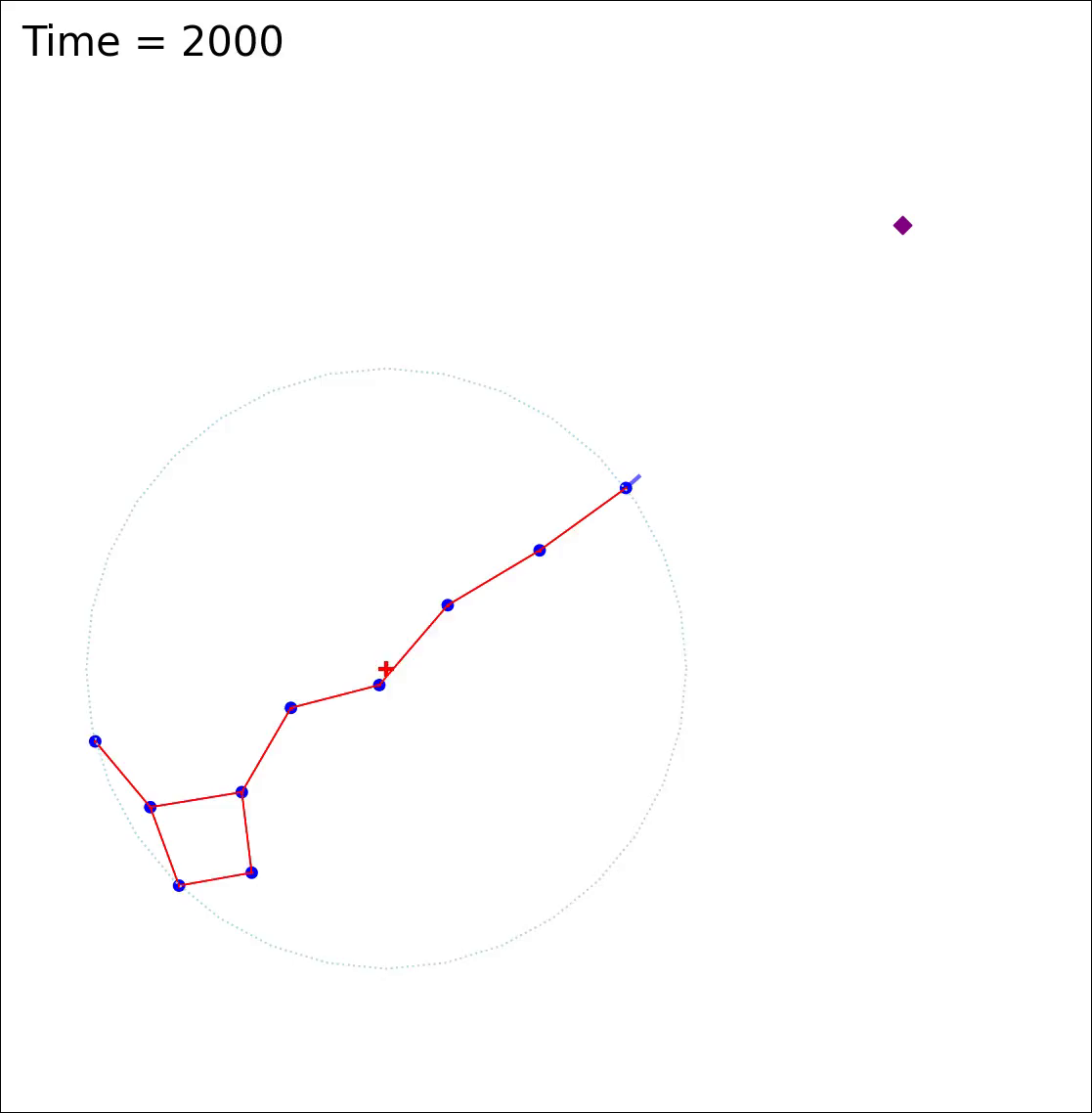}
        %\captionof{subfigure}\centering{}
    \end{minipage}
    \begin{minipage}[t]{0.30\columnwidth}
        \centering
        \includegraphics[width=\columnwidth]{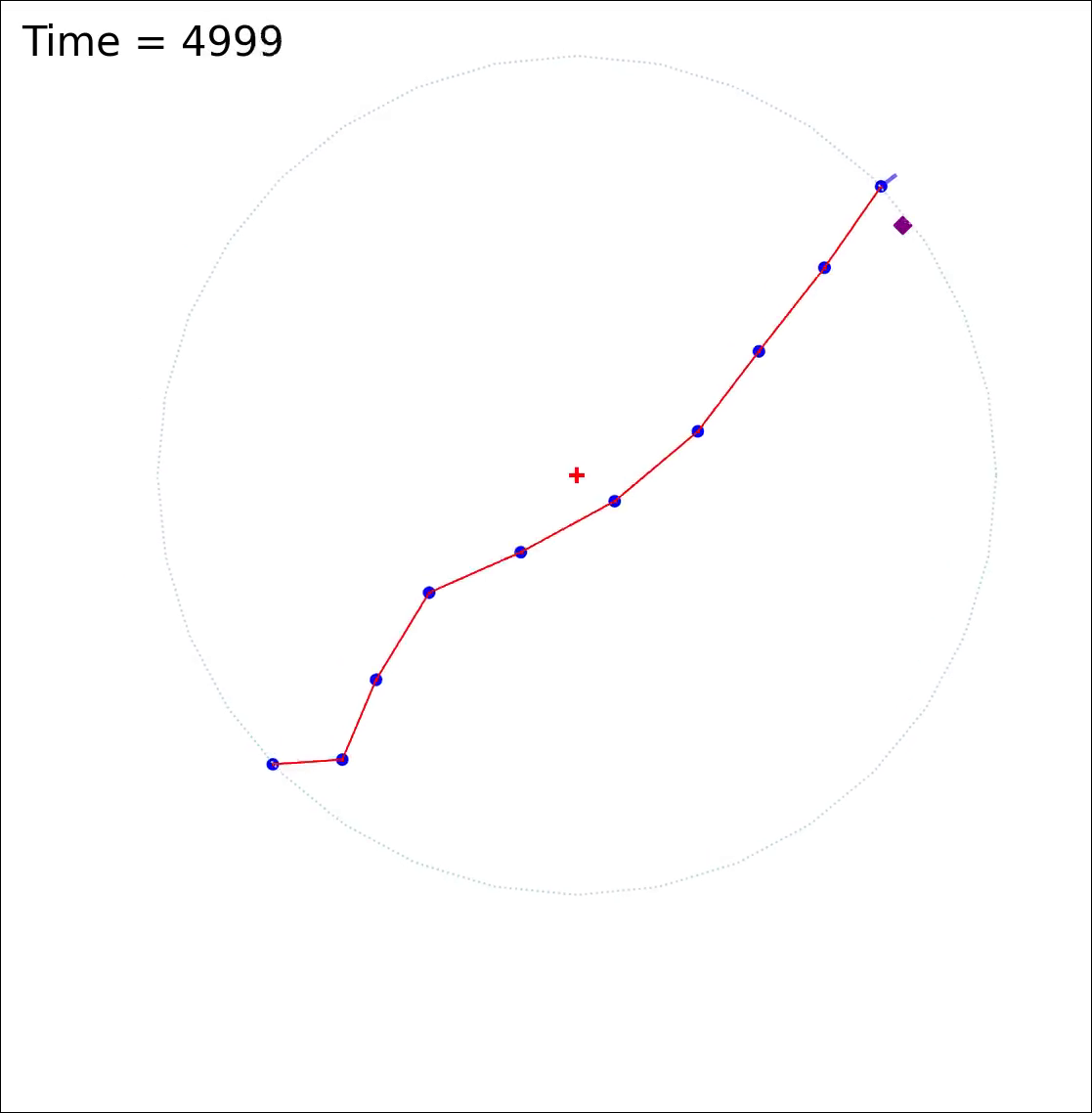}
        %\captionof{subfigure}\centering{}
    \end{minipage}
    \begin{minipage}[t]{0.3\columnwidth}
        \centering
        \includegraphics[width=\columnwidth]{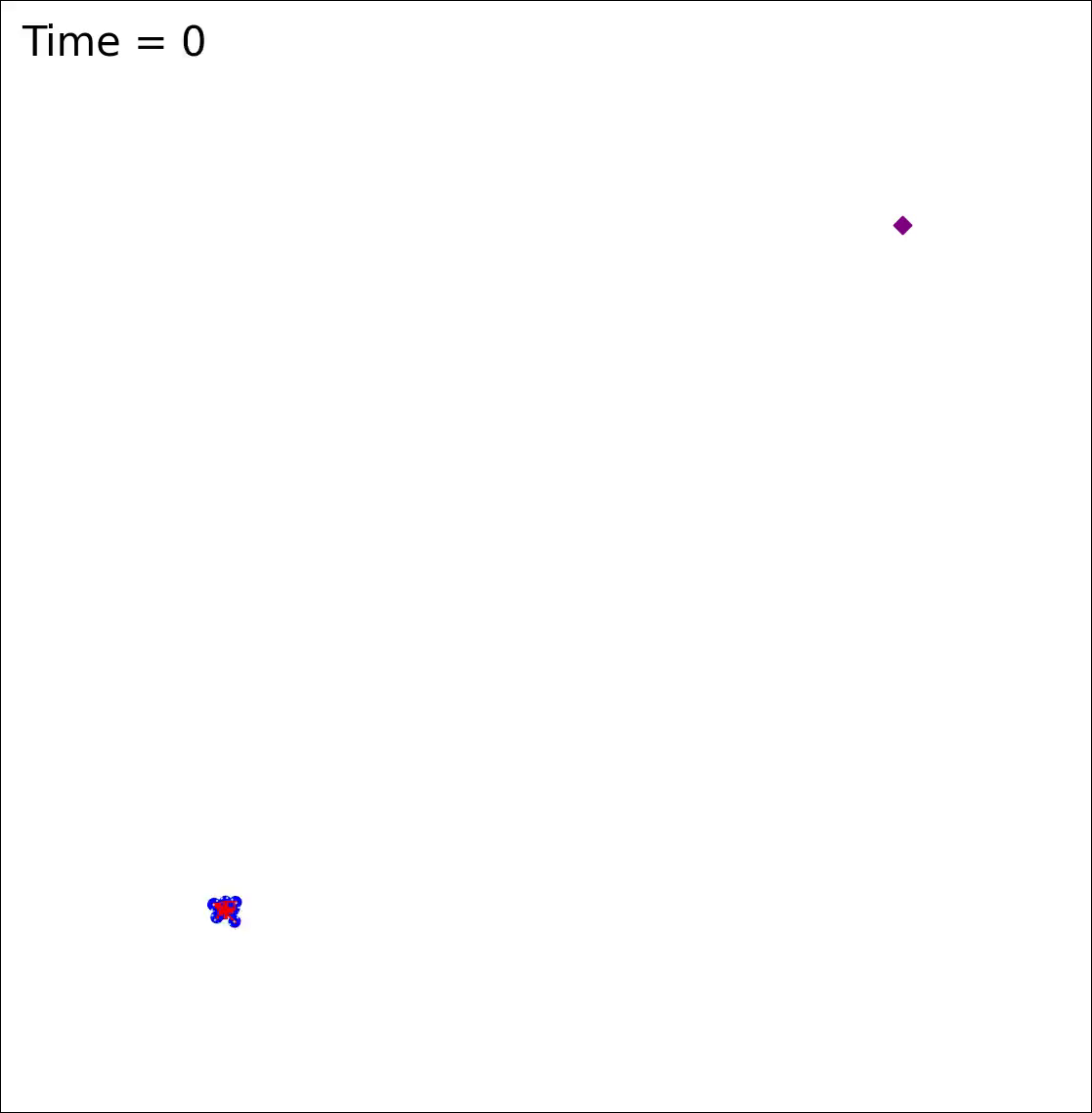}
        %\captionof{subfigure}\centering{}
        %\label{fig: snake}
    \end{minipage}
     \begin{minipage}[t]{0.3\columnwidth}
        \centering
        \includegraphics[width=\columnwidth]{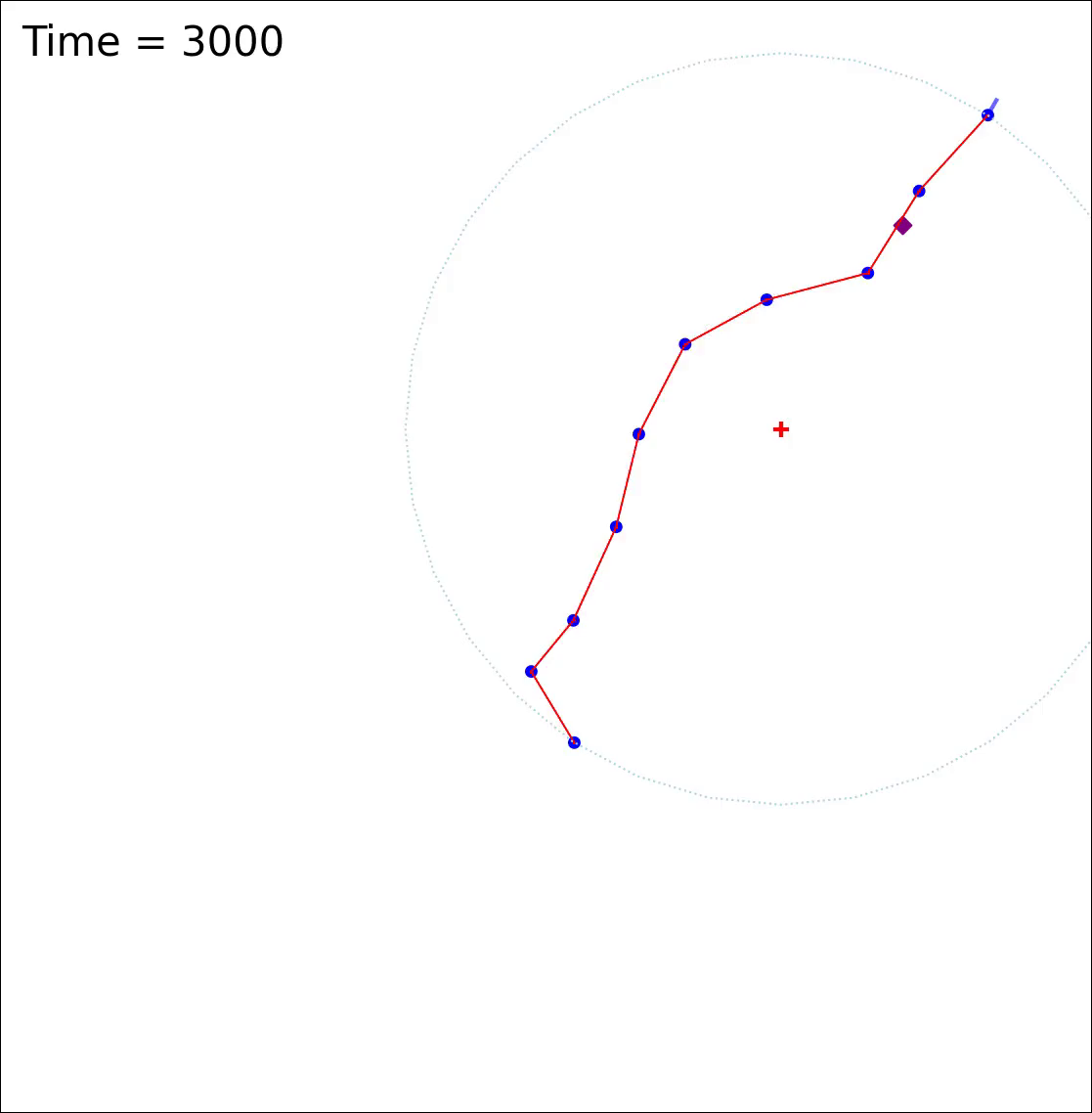}
        %\captionof{subfigure}\centering{}
    \end{minipage}
     \begin{minipage}[t]{0.3\columnwidth}
        \centering
        \includegraphics[width=\columnwidth]{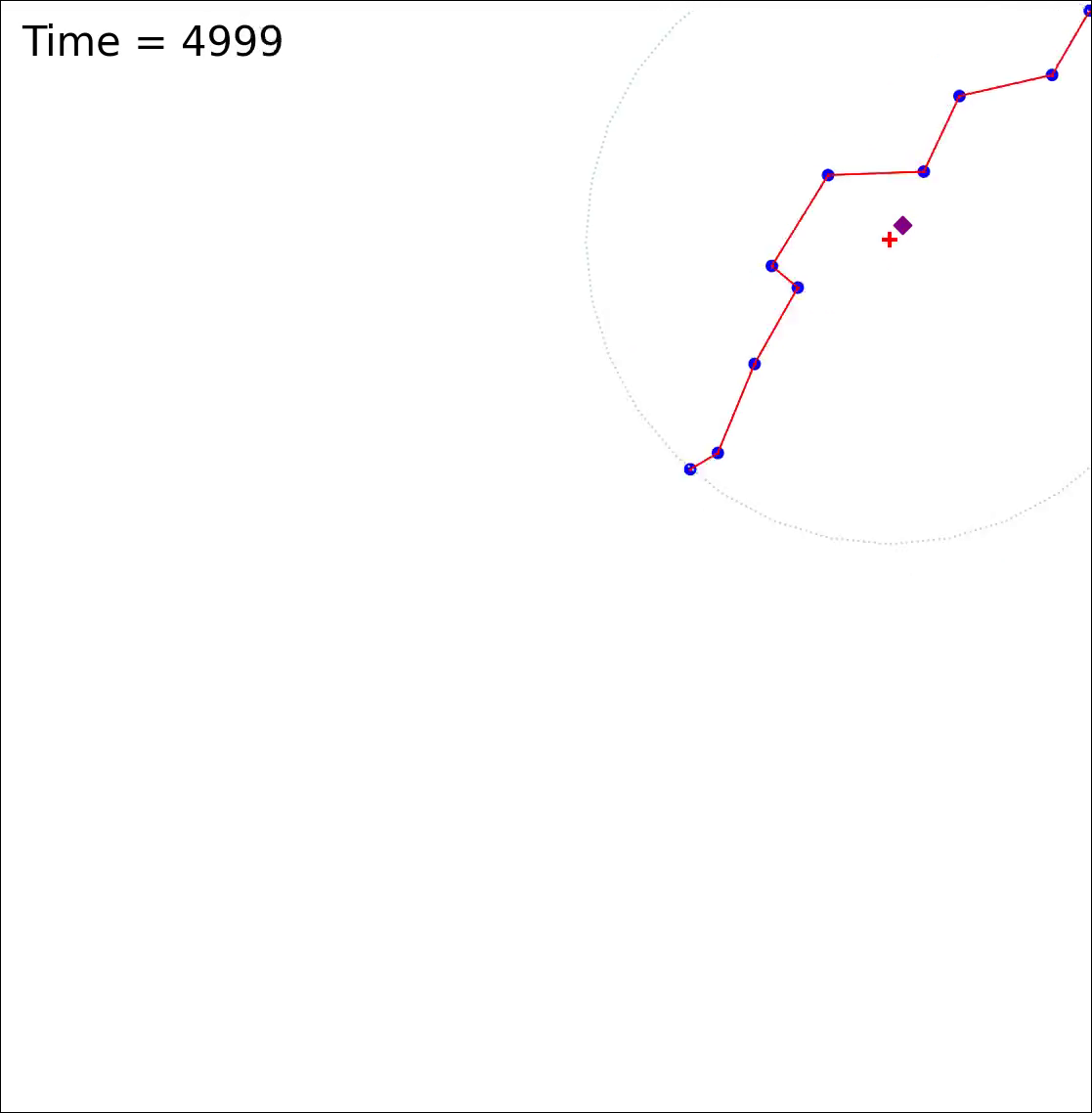}
        %\captionof{subfigure}\centering{}
    \end{minipage}\vspace{0.8mm}
     \begin{minipage}[t]{0.95\columnwidth}
        \centering
        \includegraphics[width=\columnwidth]{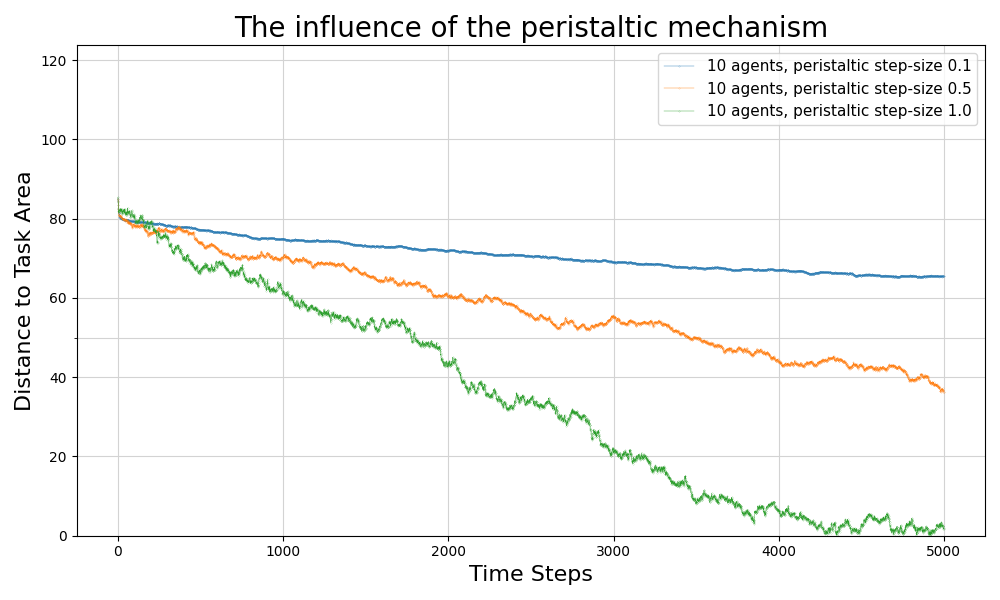}
        %\captionof{subfigure}\centering{}
    \end{minipage}
    \caption{
    The influence of the peristaltic mechanism on the swarm’s maneuverability. 
    It consists of three rows of scenario results under different peristaltic steps-sizes (top to bottom: $0.1$, $0.5$, $1.5$) over time, progressing from left to right. The swarm's location is represented by a red ``plus" sign, while the center of a task area is represented by a red diamond at the top-right side.
    The comparative graph depicts the swarm's distance to the task area under the peristaltic step-size.}
    \label{fig:Perstaltic}
\end{figure}

We tested the influence of different peristaltic step sizes on the swarm's velocity, as depicted in Fig. \ref{fig:Perstaltic} , illustrating $3$ scenario results with different step-sizes ($0.1$, $0.5$, $1.0$).  A red ``plus" sign, at the left-bottom, represents the center of the swarm's bounding circle, while a diamond at the top-right area represents the task region. In the scenario run with step size $1.0$ the swarm reached the task area, while it was far in the $0.5$ step-size scenario and even farther in the $0.1$ step-size. The graph illustrates a direct correlation between the peristaltic step size and the velocity of trail generation, however, opting for a random step size that is too large can impact velocity, as a random walk may cause agents to deviate from their intended direction towards the task area.

\subsection {Steering}
To guide the swarm towards the task area, the controller broadcasts signals containing information indicating the task area's direction relative to the swarm's location.
However, this form of communication does not guarantee delivery, and there is a probability that agents fail to receive the signals. As a result, the agents may perform a random walk rather than heading to the task area. Therefore, it is crucial to examine the effect of various reception probabilities on the swarm's mobility, especially in the case of large-scale swarms where the number of agents that don't receive the signal isn't negligible.

Having said that, in this section, we introduce a new key parameter: signal-reception probability. This parameter represents the agent's probability of receiving the transmitted signal at each time step. 

\begin{figure}[!hbt]
    \centering
    \begin{minipage}[t]{0.30\columnwidth}
        \centering
        \includegraphics[width=\columnwidth]{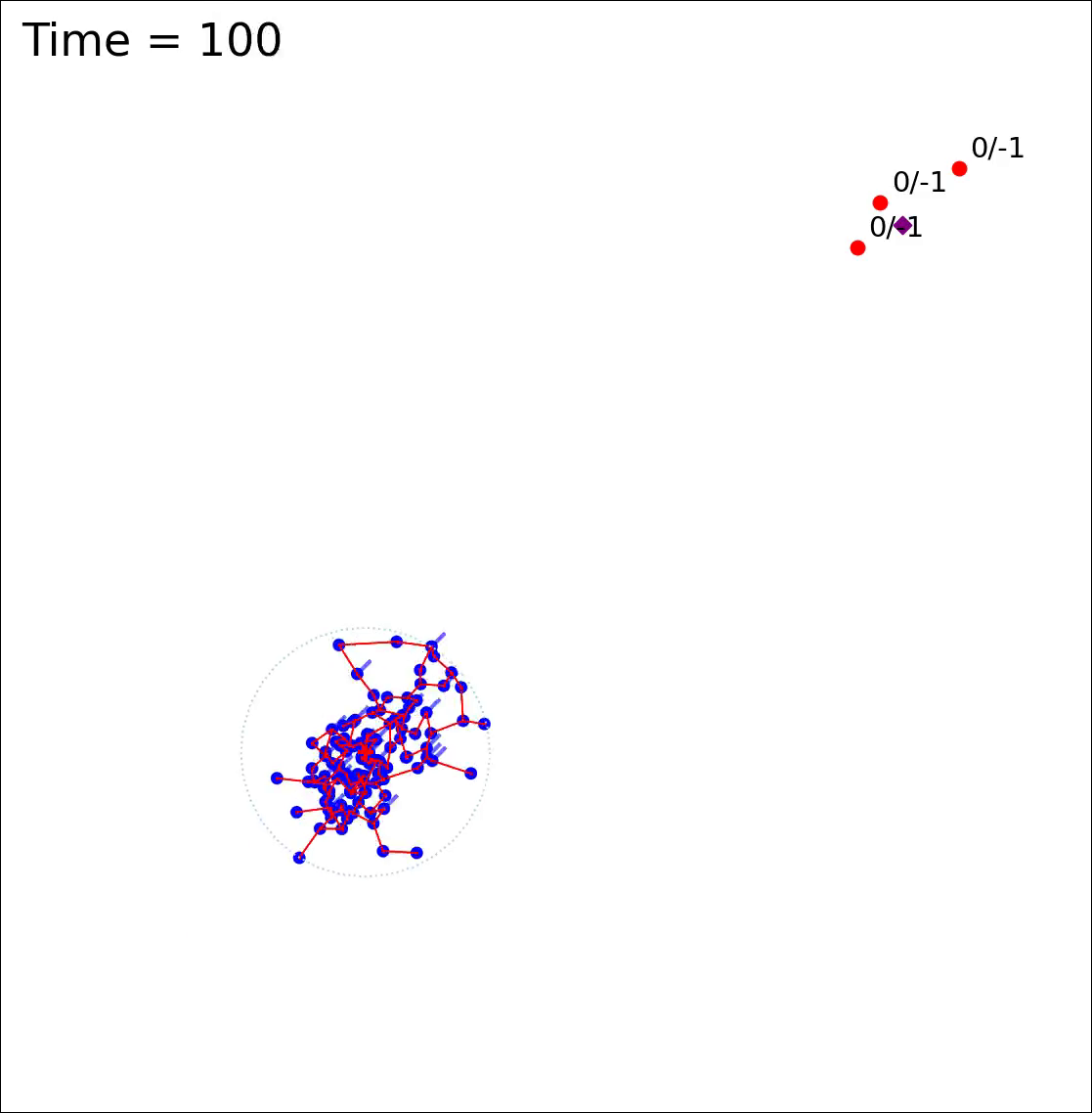}
        %\captionof{subfigure}\centering{}
    \end{minipage} \vspace{0.5mm} % Add vertical space between rows
    \begin{minipage}[t]{0.30\columnwidth}
        \centering
        \includegraphics[width=\columnwidth]{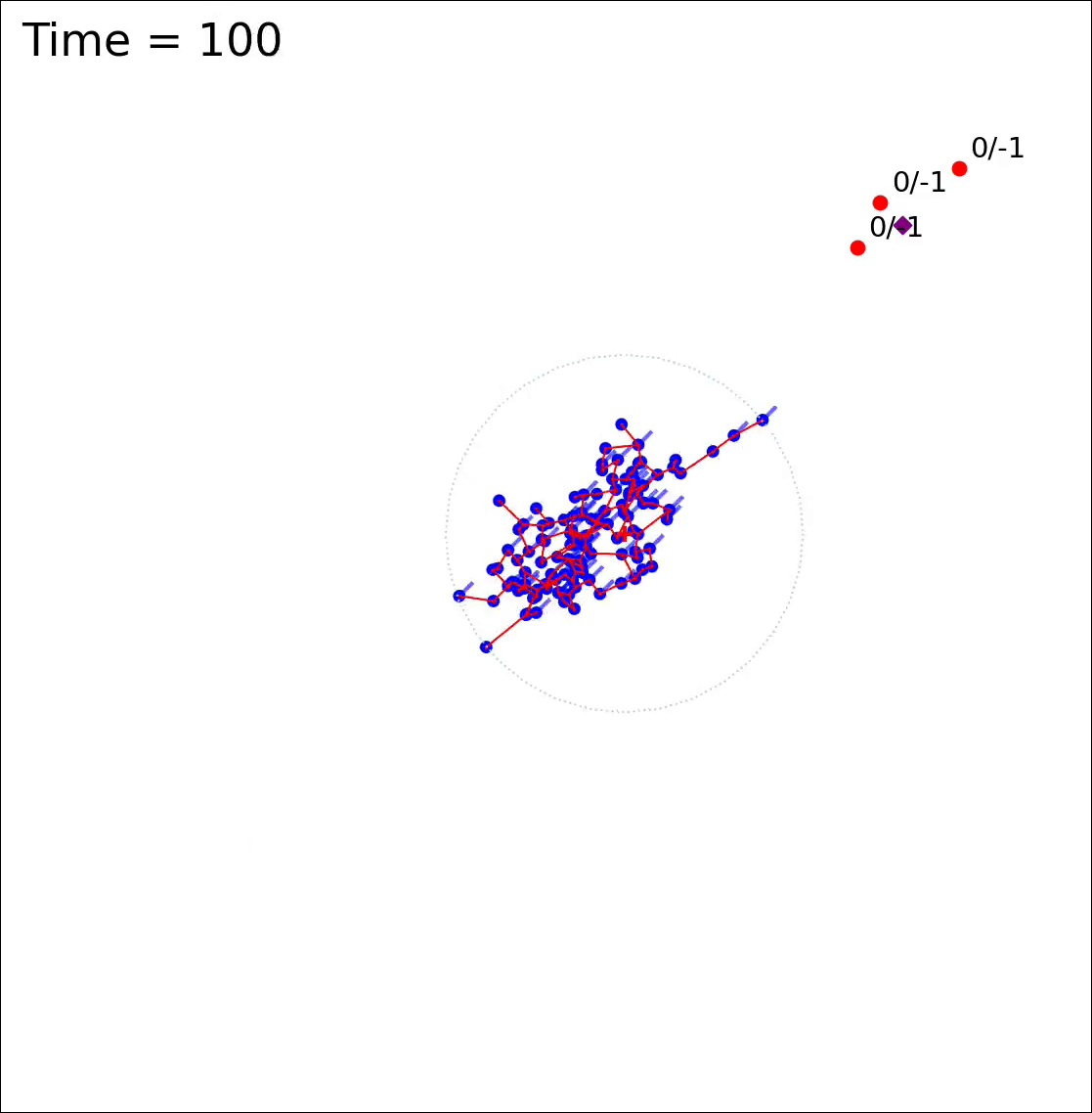}
        %\captionof{subfigure}\centering{}
    \end{minipage}
    \begin{minipage}[t]{0.30\columnwidth}
        \centering
        \includegraphics[width=\columnwidth]{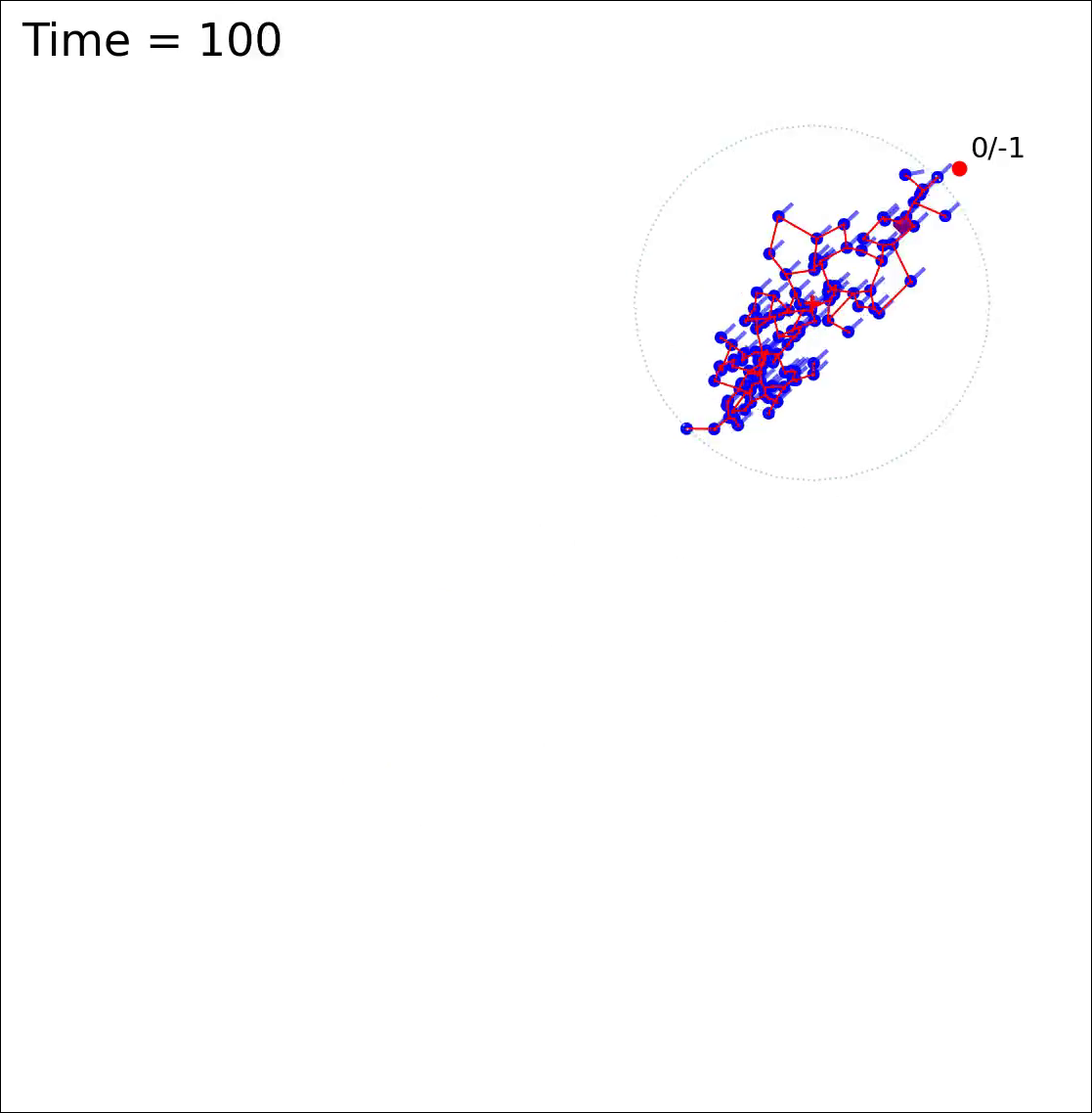}
        %\captionof{subfigure}\centering{}
    \end{minipage}
    \begin{minipage}[t]{0.30\columnwidth}
        \centering
        \includegraphics[width=\columnwidth]{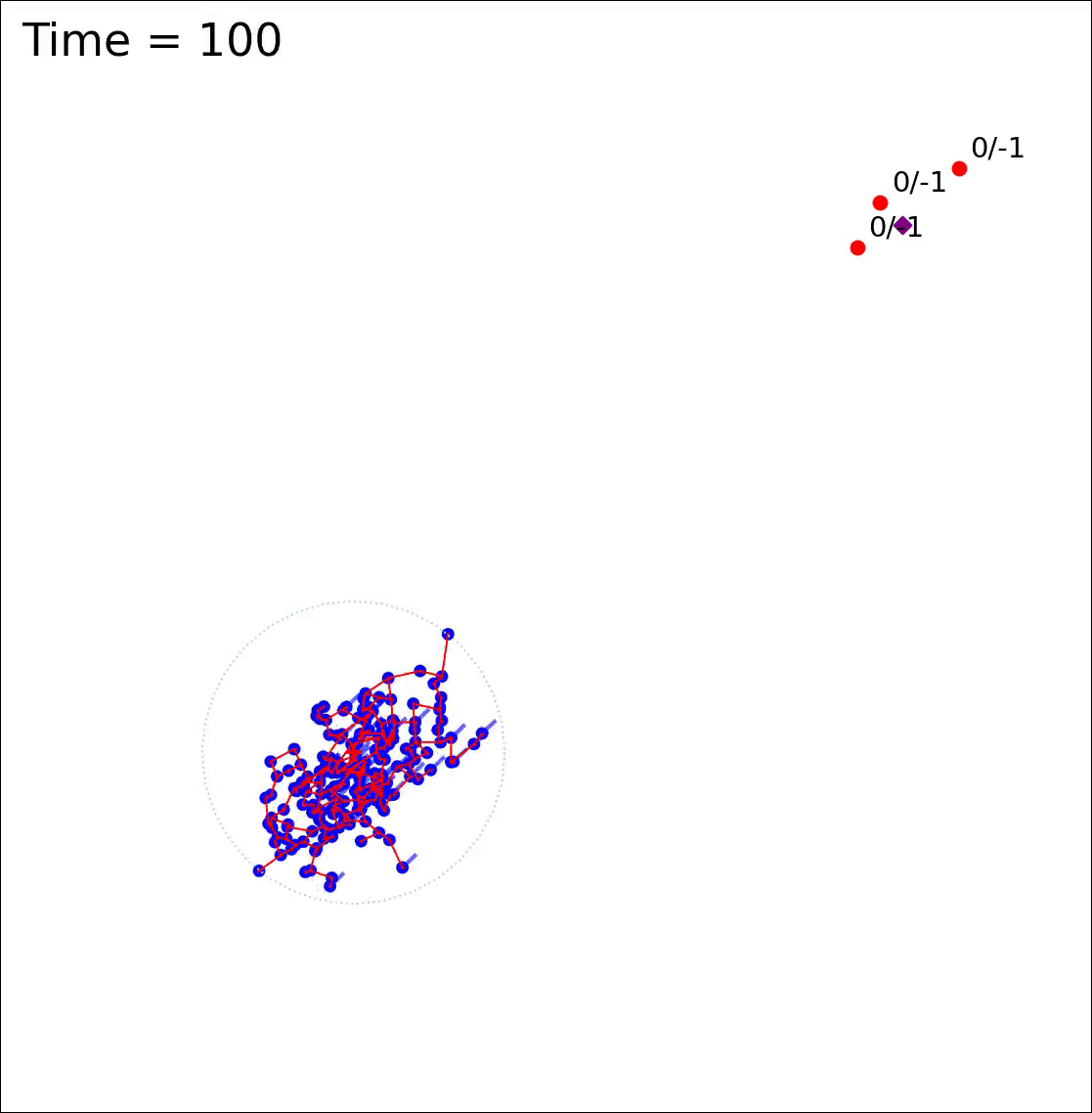}
        %\captionof{subfigure}\centering{}
    \end{minipage}
    \begin{minipage}[t]{0.30\columnwidth}
        \centering
        \includegraphics[width=\columnwidth]{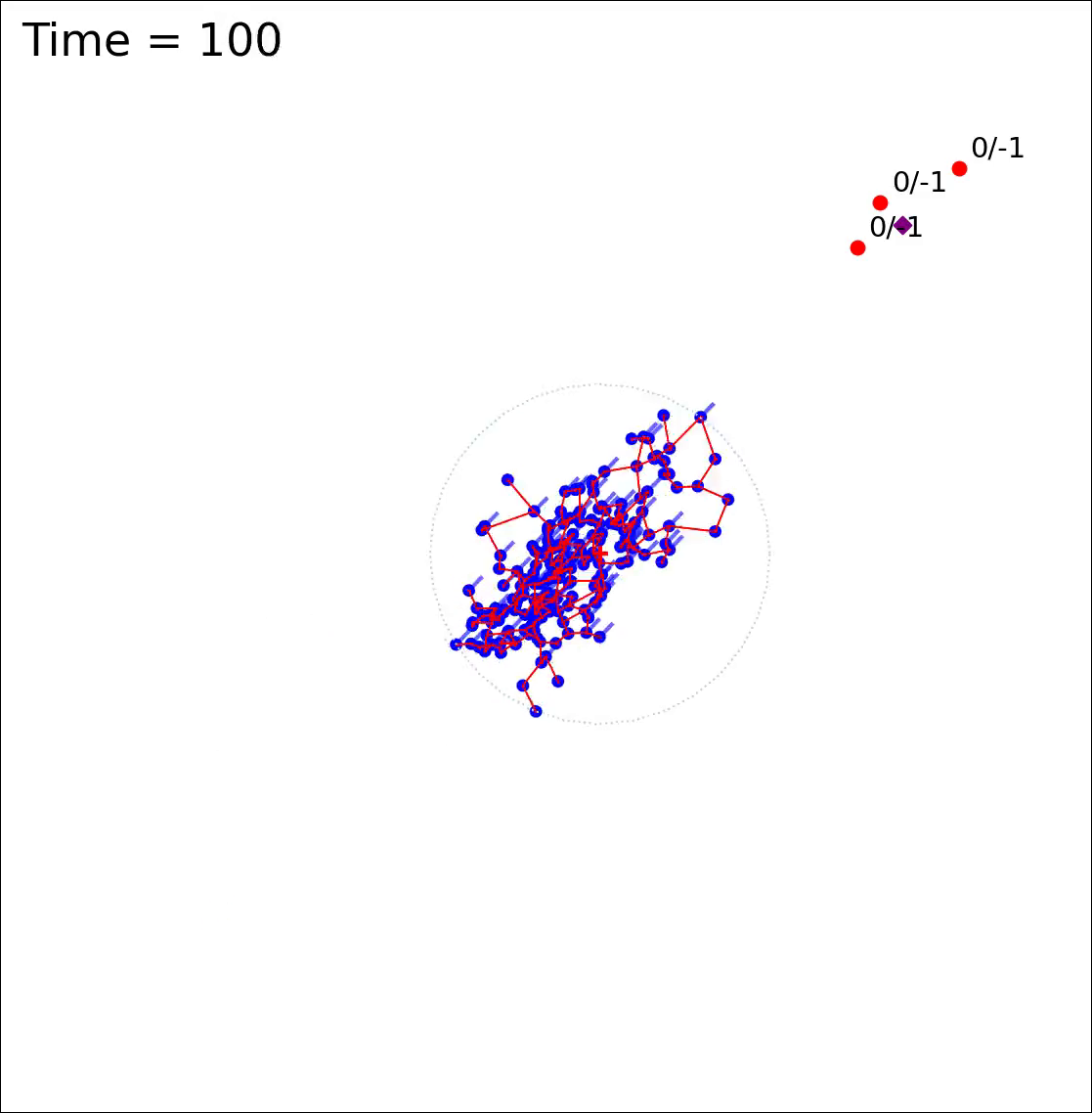}
        %\captionof{subfigure}\centering{}
    \end{minipage}
    \begin{minipage}[t]{0.30\columnwidth}
        \centering
        \includegraphics[width=\columnwidth]{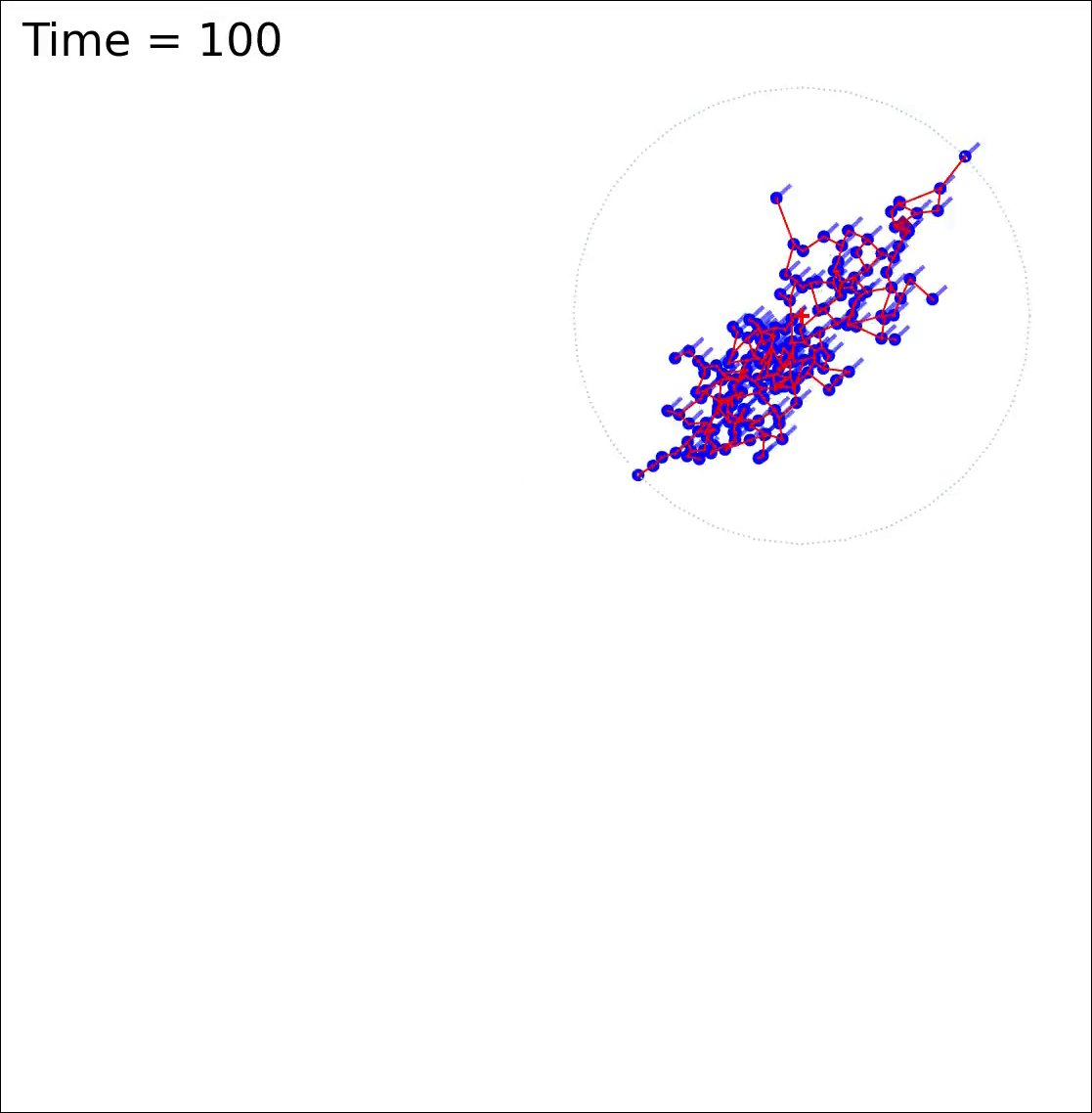}
        %\captionof{subfigure}\centering{}
        %\label{fig: snake}
    \end{minipage}\vspace{0.8mm}
    \begin{minipage}[t]{0.95\columnwidth}
        \centering
        \includegraphics[width=\columnwidth]{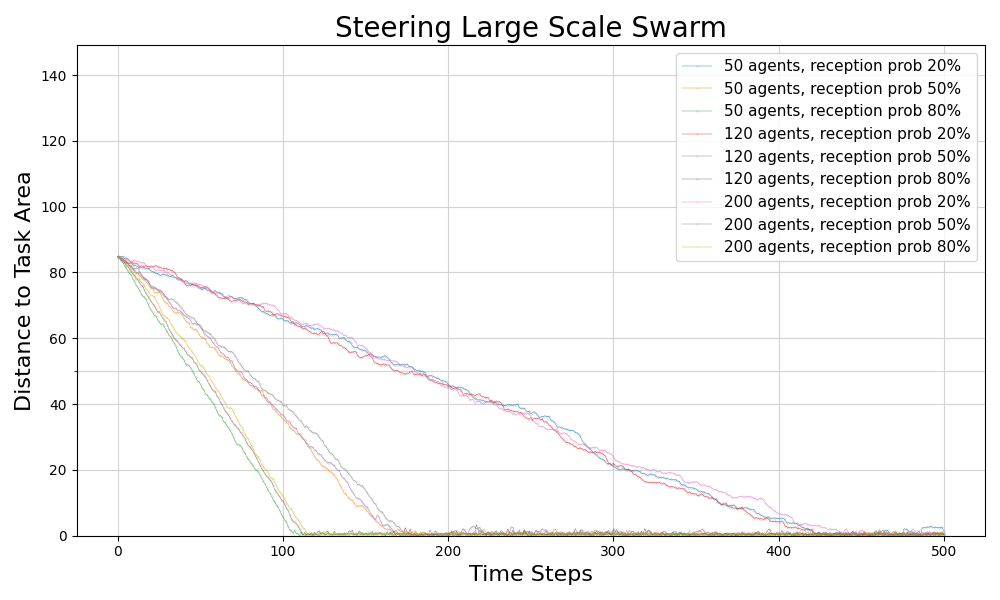}
        %\captionof{subfigure}\centering{}
        %\label{fig: snake}
    \end{minipage}
    \caption{
    The influence of various reception probabilities and swarm sizes on the swarm mobility towards the task area. The figure shows two rows of images taken at time=$100$. The top row, from left to right, displays 120 agents' positions with a reception probability of $20\%$, $50\%$, and $80\%$. The bottom row from, left to right, displays 200 agents with a reception probability of $20\%$, $50\%$, and $80\%$ from right to left.
    The comparative graph depicts distance to task area of $50$, $120$, $200$ agents with reception probability of $20\%$, $50\%$, and $80\%$.}
    \label{fig: Steering Large Scale Swarm}
\end{figure}

The results of nine scenario executions are depicted in Fig \ref{fig: Steering Large Scale Swarm}. The scenarios consist of 3 differently-sized guided swarms (with $50$, $120$, and $200$ agents), with various reception probabilities ($20\%$, $50\%$, and $80\%$.
The top two rows (samples of $120$ and $200$ agents) taken at $Time step = 100$ provide a course indication that motion toward the tasks is primarily influenced by the reception probability. The comparative graph of the swarms further emphasizes this phenomenon, illustrating that the swarm's velocity is proportional to the agent's reception probability. This implies that the ratio between the anonymous leaders to the swarm's size has a critical influence.\\

\section {Conclusion}
This paper introduces a bio-mimetic model for a distributed control of swarms, inspired by observations of natural swarm behaviour. The model consists of three distinct layers: agent, formation, and global swarm, behaviour promoting cohesiveness, exploration, task allocation, and maneuverability.

Consistent with the paradigms mentioned earlier, the algorithms enable a limited sensory swarm to scan and locate tasks, while maintaining connectivity. Notably, all these operations occur through distributed control, without explicit inter-agent coordination.

We also introduced a steering model with an observer that broadcasts guiding signals, creating ``Anonymous Leaders" that motivate a collective movement of a swarm toward the task area without centralized control and direct inter-agent communication. We examined the influence of reception probability on the swarm movement, highlighting its importance while presenting the agnostic nature of the swarm to its size.

Our hypotheses were simulated and tested across various scenarios. The outcome results highlight the significance of the key parameters and their influence on swarm behavior.
This research contributes to swarm control applications and improves autonomous behaviour, especially in challenging conditions and terrains such as those encountered in search and rescue missions. 
Future research directions in this field could focus on coping with objects or obstacles located in the environment that impede agents from reaching their tasks.

%%%%%%%%%%%%%%%%% BIBLIOGRAPHY IN THE LaTeX file !!!!! %%%%%%%%%%%%%%%%%%%%%%

\bibliographystyle{unsrt}
\bibliography{References}

\end{document}